\documentclass[twocolumn, aps, prb, reprint, letterpaper, notitlepage, superscriptaddress]{revtex4-2}
\usepackage{amsmath,amssymb}
\usepackage{graphicx}
\usepackage{physics}
\usepackage{hyperref}
\usepackage{xcolor}

\newcommand{\NV}{\text{\tiny NV}}

\newcommand{\E}{\text{\tiny E}}
\newcommand{\RE}{\text{\tiny R-E}}
\newcommand{\CP}{\text{\tiny CP}}
\newcommand{\X}{\text{\tiny X}}

\begin{document}
\title{Self-consistent Noise Characterization of Quantum Devices}

\author{Won Kyu Calvin Sun}
\affiliation{Research Laboratory of Electronics, Massachusetts Institute of Technology, Cambridge, Massachusetts 02139, USA}
\affiliation{Department of Nuclear Science and Engineering, Massachusetts Institute of Technology, Cambridge, MA 02139, USA }

\author{Paola Cappellaro}\email{pcappell@mit.edu}
\affiliation{Research Laboratory of Electronics, Massachusetts Institute of Technology, Cambridge, Massachusetts 02139, USA}
\affiliation{Department of Nuclear Science and Engineering, Massachusetts Institute of Technology, Cambridge, MA 02139, USA }
\affiliation{Department of Physics, Massachusetts Institute of Technology, Cambridge, MA 02139, USA }

\date{\today}

\begin{abstract}
Characterizing and understanding the environment affecting quantum systems is critical to elucidate its physical properties and engineer better quantum devices.
We develop an approach to reduce the quantum environment causing single-qubit dephasing to a simple yet predictive noise model. Our approach, inspired by quantum noise spectroscopy, is to define a `self-consistent’ classical  noise spectrum, that is, compatible with all observed decoherence under various qubit dynamics.
We demonstrate the power and limits of our approach by characterizing, with nanoscale spatial resolution, the noise experienced by  two  electronic spins in diamond that, despite their proximity, surprisingly reveal the presence of a complex quantum spin environment, both classically-reducible and not.
Our results overcome the limitations of existing noise spectroscopy methods, and highlight the importance of finding predictive models to accurately characterize the underlying environment. Extending our work to multiqubit systems would enable spatially-resolved quantum sensing of complex environments and quantum device characterization, notably to identify correlated noise between qubits, which is crucial for practical realization of quantum error correction.\end{abstract}

\maketitle


\section{Introduction}\label{sec:intro}
The performance of quantum devices is often limited by the effects of their environment, even if the environment could be tamed or even turned into a resource if it could be properly characterized~\cite{Goldstein11,Taminiau12,Abobeih19a,Bradley19,Liu19,Cooper20,Jackson21,Ruskuc22}.
Unfortunately, a full characterization of the environment is usually not possible and one has to rely on a simplified model of the noise sources.
For simpler quantum systems such as qubits and qutrits, it is in principle always possible  to reduce a complex quantum environment to a classical noise (spectrum) model, at least for a fixed dynamics of the total system~\cite{Crow14,Helm11,Helm09}.
However, this noise model is not guaranteed to be predictive when the system (or bath) dynamics is changed by control, as is the case for quantum devices.
Obtaining a classical noise spectrum that can describe the system dynamics under a broad set of controls and predict its performance would be highly desirable, not only to enable practical characterization of unknown complex many-body environments (e.g., for applications in quantum sensing or quantum device characterization), but also to engineer more robust quantum devices and control sequences tailored to the noise.

\begin{figure}
\centering
\includegraphics[width=\columnwidth]{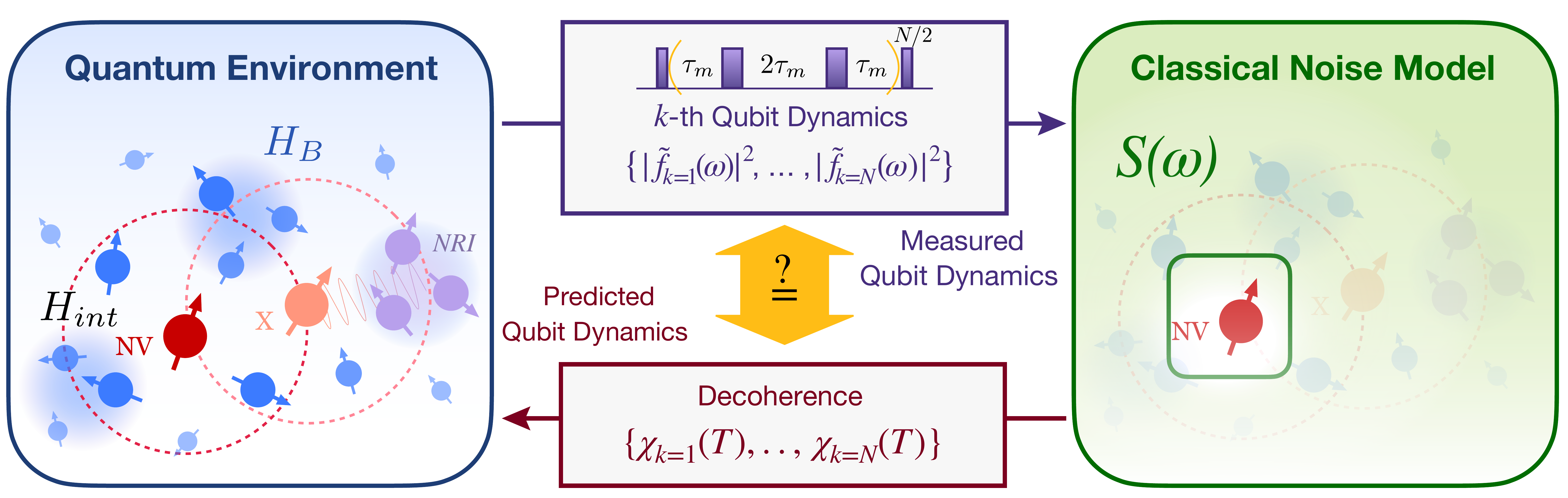}
\caption{
\textbf{Reducing a quantum environment to a self-consistent classical noise model.}
To model a quantum environment, we attempt to develop a classical noise model $S(\omega)$ that is consistent with the set of all observed decoherence under various controlled dynamics. {When such a `self-consistent’ noise model is possible, as demonstrated in this paper experimentally for an NV electronic spin in diamond but not a nearby interacting electronic spin X several nanometers away, we further verify that the self-consistent model has predictive power even under new dynamics, confirming that it accurately models the underlying quantum bath.}
}
\label{FigScheme}
\end{figure}

In this paper, we demonstrate an approach to build a practical yet predictive noise model of qubit decoherence.
Our approach is to form a `self-consistent' classical noise model --- that is, consistent with all observed decoherence under various qubit dynamics --- by reconciling complementary approaches to noise spectroscopy.
Crucially, by reconciling limitations of existing methods, we demonstrate that it succeeds even when {the existing methods fail} to yield the correct noise model, and is further able to predict the system dynamics under additional control sequences.
If such a self-consistent noise model is possible, this indicates that the underlying (quantum) bath can be effectively reduced to a classical Gaussian noise process, enabling practical characterization of the bath with predictive power. We demonstrate this experimentally, by building a self-consistent noise model of the electronic spin of a nitrogen-vacancy (NV) center in diamond and subsequently verify that it is predictive even under new qubit dynamics.
On the other hand, if a self-consistent model is not possible, this indicates that the underlying bath is sufficiently complex, either of quantum or of non-Gaussian nature. We verify this experimentally with another electronic spin near the NV --- and indeed with further investigation verify the quantum nature of its local environment.
Finally, having characterized the bath of two nearby electronic spins in diamond, we are able to probe, with nanoscale spatial resolution, the dominant source of noise common to both qubits arising from the quasistatic many-body electronic spin bath. The noise model reveals the local spin density and timescale of spin bath dynamics with nanoscale variations, information which is inaccessible by conventional nuclear magnetic resonance (NMR) or ensemble-sensor techniques.

\begin{figure*}
\centering
\includegraphics[width=2\columnwidth]{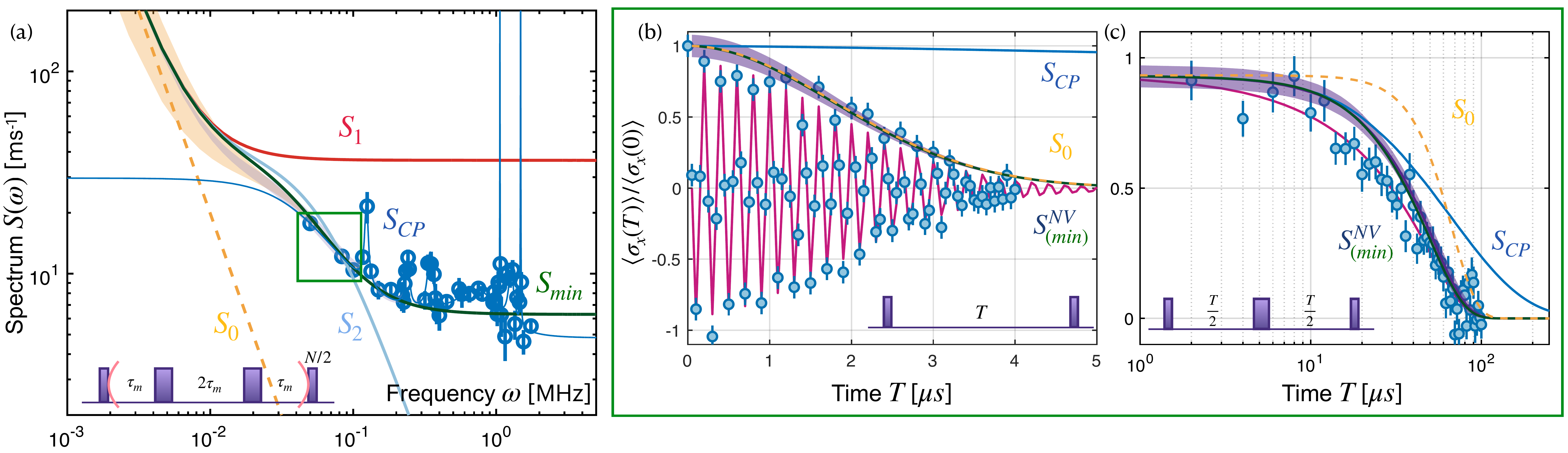}
\caption{
\textbf{Self-consistent noise model of an NV electronic spin in diamond.}
The (minimal) self-consistent noise model $S^{\NV}_{(\text{min})}(\omega)$ is presented, along with noise model candidates $S_{1,2}$ (consistent with R-E but not CPMG dynamics) and $S_{\CP}$ (vice versa).
While as shown both R-E- and CPMG-based methods fail to yield the correct noise model due to their limitations, by reconciling them our method succeeds.
{(a)} The noise models are shown against the measured decoherence (markers) under multiple CPMG dynamics, $S(\omega_m)$. Note that $S_{1,2}$ fail to be predictive under higher-frequency noise.
{(b,c)} Decay under Ramsey (b) and spin echo (c) dynamics is measured (blue circles) and fitted (red curve) to perform R-E-noise spectroscopy.
Note that $S_{\CP}$ fails to be predictive under Ramsey or echo dynamics (zero or low-frequency noise).
The controlled qubit dynamics (pulse sequence) is shown below; green boxes indicate the minimal experimental measurements used to inform $S^{\NV}_{\text{min}}(\omega)$, which is further predictive of new dynamics (see Fig.~\ref{FigPredictivePower}). The left green box in (a) contains $S_{\CP}(\omega_m)=\{17.5,~12,~10.5\} ~\text{ms}^{-1}$ at $\omega_m=(2\pi)\{0.05,~0.08\bar{3},~0.10\}~\text{MHz}$, respectively.
}
\label{FigSCNoiseModel}
\end{figure*}

\section{Quantum Noise Spectroscopy}\label{sec:bkg}
Several protocols for noise spectroscopy have been developed thus far, ranging from simple sequences~\cite{Dobrovitski08,deLange10,deLange12} to more complex continuous~\cite{Yan13,Wang20,Wang21} and pulsed~\cite{Cywinski08,Alvarez11,Yuge11,Ryan10} control.
They have successfully elucidated noise sources (from local fluctuators~\cite{Paladino02, Galperin06, Bergli07, Cywinski08,Chen18}
to spin environments~\cite{Dobrovitski08,deLange10,deLange12,Alvarez11,Yuge11,Wang13}),
and their accuracy to reproduce a given classical noise has been evaluated~\cite{Szankowski18}.
However, much less attention has been paid to analyze their predictive power especially when the reconstructed noise spectrum is only an approximation to the real noise, i.e., whether because it arises from a quantum system~\cite{Hernandez-Gomez18} or a complex classical source~\cite{Norris16,Kwiatkowski18,Sung19}---or more simply due to experimental limitations.
Here, to achieve a predictive noise model, we propose to build a self-consistent noise spectrum by combining complementary approaches.

The simplest approach, which we call R-E-noise spectroscopy, utilizes only decoherence under the free evolution [Ramsey, (R)] and spin echo (E) experiments. The knowledge of their decay functionals and decay times $T_2^*$ (R) and $T_2$ (E) may be sufficient to fully characterize a noise model $S(\omega|\vec p)$ with unknown model parameters $\vec p$~\cite{SOM}.
While minimal in experimental cost, this method requires a noise model that is already known and sufficiently simple to uniquely identify $\vec p$~\cite{Dobrovitski08,deLange10,deLange12}. Furthermore, it can only investigate low-frequency noise ($\omega<T_2^{-1}$).

A more general approach based on dynamical-decoupling sequences with equidistant $\pi$ pulses [Carr-Purcell-Meiboom-Gill (CPMG) pulse sequences] can in principle reconstruct the full noise spectrum. Under the filter-function formalism, each CPMG experiment of inter-pulse length $2\tau_m$ forms a filter $|\tilde{f}_T(\omega)|^2$ that approximates a delta function $\delta(\omega\!-\!\omega_m),$ $\omega_m=(2\pi)(4\tau_m)^{-1}$. This allows direct measurement of $S(\omega_m)$ from the simple-exponential decay $\chi_m(T)$ under CPMG pulse sequences, where
\begin{align}
\label{eq:chiCPm}
	\chi_m(T) &= \frac{1}{2} \int S(\omega)|\tilde{f}_T(\omega)|^2\frac{d\omega}{2\pi}  \approx  \frac{4}{\pi^2} S(\omega_m) T.
\end{align}
While this method can characterize arbitrary, unknown noise spectra with high-resolution, it comes at increased experimental cost, as one CPMG experiment is needed per frequency.
Furthermore, the bandwidth, while much broader, is still bounded by the coherence time $T_2$ and Rabi frequency $\Omega_0$, $T_2^{-1} < \omega_m \ll \Omega_0$~\cite{Yuge11}. In particular, low frequencies are harder to reach in the presence of strong noise.

\section{Self-consistent Noise Characterization}\label{sec:experiment}
Combining these techniques, we demonstrate how to obtain a self-consistent classical model. We start with a minimal noise model, consistent with initial experimental data, and incrementally refine it as necessary to be consistent with additional experiments. While other strategies are possible, this minimizes the experimental cost. We first demonstrate the protocol in the concrete case of an NV center in diamond (Fig.~\ref{FigScheme}).

\subsection{NV electronic spin qubit}\label{sec:NV}
The first step is to measure the NV Ramsey dynamics.
We used the $m_s=\{0,-1\}$ states of the NV electronic spin (electronic spin $S=1$) in an external static magnetic field of strength $B_0\approx 350~\text{G}$ aligned approximately along the NV axis. The control was achieved with a single-tone, resonant microwave of $\Omega_0^\NV\approx 6.9~\text{MHz}$ amplitude to drive both $^{15}$NV hyperfine transitions ($A_{zz}\approx 3.2~\text{MHz}$).

Observing a Gaussian decay under Ramsey control [Fig.~\ref{FigSCNoiseModel}(b)], we assume as our minimal model an Ornstein-Uhlenbeck (OU) process
\begin{equation}\label{eq:OU}
	S(\omega|b,\tau_c)  =  \frac{b^2 (2\tau_c)}{1+(\omega \tau_c)^2},
\end{equation}
characterized by two parameters $(b,\tau_c)$.
Indeed, a quasistatic or ``slow'' OU noise, $(b_s\tau_s)\!\gg \! 1$,  predicts a Gaussian decay, $\chi_R(T)=(b_s T)^2/2\equiv(T/T_2^*)^2$.
More generally, the slow-OU noise has successfully modeled noise from a slowly fluctuating spin bath~\cite{deLange10,deLange12,Wang13} and is expected~\cite{Cooper20} to be the dominant noise in our system
\footnote{
We remark that a simpler noise model candidate $S_0\! \propto \!\delta(\omega)$, which also yields a gaussian Ramsey decay, has been ruled out as inconsistent with the echo.}.
Then, fitting for $T_2^*$ we identify one of two unknown parameters, $b_s=0.56(2)~\text{MHz}$.

Given a working model $S_0\!=\!S_s$ consistent with Ramsey dynamics, we can ask whether it is {already} predictive of echo dynamics.
Unfortunately, {we find that it is not, as} while $S_0$ predicts a stretched-exponential $\chi_{\E}(T)\approx (b_s^2T^3)/(12\tau_s) \equiv (T/T_2)^3$, the NV echo is dominantly simple exponential [Fig.~\ref{FigSCNoiseModel}(c)].
{Note that similarly we could have started with the knowledge of NV echo decay to first search for a minimal (single-termed) noise model consistent with echo dynamics and test whether it is predictive of Ramsey dynamics.} In such a case, we would arrive at either a fast-OU noise $S_f$ ($\tau_f\!\ll\!  T$) or white-noise $S_w$, which both yield an exponential decay. However, neither are consistent with NV Ramsey dynamics.

{This suggests that the environment around the NV is sufficiently complex so as not to be reduced to a single independent noise process.}
We thus introduce minimal complexity to the working model by considering two terms and immediately find two valid models: a single-OU plus white-noise model $S_1 = S_s\!+\!S_w$, and a double-OU model $S_2= S_s\!+\!S_f$. Both $S_{1}$ and $S_{2}$ predict the same competing decay under echo dynamics with two characteristic timescales: ${\chi}_{\E}(T) = (T/T_2)^3 + T/T_0$,
where $T_2=(12\tau_s/b_s^2)^{1/3}$ and $T_0 = 2/S_w ~(\text{for}~S_1)$ or $T_0=(b_f^2\tau_f)^{-1} ~(\text{for}~ S_2)$.
{In fact,} fitting the NV echo to this more complex $S_{1,2}$ yields the best fit versus the simpler models with a single characteristic decay, confirming their validity.
{Notably,} a similar multicomponent bath model has successfully described the noise of shallow NVs~\cite{Myers14,Romach15}, with a slow bath typical of bulk NVs accompanied by a faster bath due to paramagnetic centers on the surface.
We can further identify some of the remaining unknowns, with $T_2=69(6)~\mu\text{s}$ (hence $\tau_s\!=\!8(2)~\text{ms}$) and $T_0=55(8)~\mu\text{s}$ [hence $S_w = 36(5)~\text{kHz}$].

Having completed R-E-noise spectroscopy, its three main limitations are observed~\cite{SOM}:
{(i) It is in general insufficient to characterize arbitrary noise models (e.g., here $S_1$ with three unknown model parameters could be fully characterized while $S_2$ with four unknowns could not). (ii) It cannot help identify which noise model is the true (or at least more accurate) noise model as it cannot discriminate between models predicting the same time-domain decay functionals (e.g., while $S_{1,2}$ are spectrally distinct, they predict the same decay under R and E dynamics).}
(iii) Furthermore, it is oblivious to noise at higher frequencies $\omega > T_2^{-1}$. To address these limitations, we turn to CPMG-based noise spectroscopy.

To achieve with minimal experimental cost a self-consistent noise model {$S^{}_{\text{min}}$} predictive of Ramsey, echo, and CPMG dynamics, the first step is to simply check whether any working model $S_{\RE}$ is already predictive of CPMG. This can be done by solving and checking
\begin{equation}\label{eq:checkForConsistency}
  S_{\CP}(\omega_m)  \stackrel{(?)}{=} S_{\RE}(\omega\!=\!\omega_m),
\end{equation}
where the left-hand side is given by experimental CPMG measurements at $\omega_m=(2\pi)(4\tau_m)^{-1}$ and the right-hand side is given by the {candidate} model evaluated at $\omega\!=\!\omega_m$.
Therefore, given a model with $q$ remaining unknown parameters, we need $(q\!+\!1)$ measurements (equations) to verify whether the model is self-consistent:  the first $q$ equations to solve for the $q$ unknowns---thereby identifying all model parameters $\vec p$ of $S_{\RE}(\omega|\vec{p})$---and the last measurement to check whether the model is predictive of a new CPMG experiment at $\omega_{q+1}$.

We apply this protocol to candidate models $S_{1,2}$, utilizing (up to) three CPMG experiments [Fig.~\ref{FigSCNoiseModel}(a)].
$S_1$, with $q\!=\!0$ unknowns, can be immediately checked.
As seen in Fig.~\ref{FigSCNoiseModel}(a), the significant relative error $\epsilon=[S_{\CP}(\omega_{q+1})\!-\!S_{1}(\omega_{q+1})]/S_{\CP}(\omega_{q+1})>1$ rules out $S_1$.
$S_2$, with $q\!=\!1$, must first be characterized by solving one equation.
This yields a unique solution $(b_f,\tau_f)\approx(74~\text{kHz},~3.3~\mu \text{s})$, suggesting validity of $S_2$.
However, it predicts with a small yet statistically significant error $\epsilon\!=\!0.24\!-\!0.38$ at higher frequencies [Fig.~\ref{FigSCNoiseModel}(a)].
Therefore, to improve upon the working model we again introduce minimal complexity, to include a small white-noise term {$S_w$ which is consistent with all observed dynamics thus far}, yielding $S^{\NV}_{\text{min}}\!=\! S_s+S_f+S_w$.
As this model has $q\!=\!2$ unknowns, we require three measurements to check for self-consistency.
This yields a unique $(b_f,\tau_f,S_w)\approx(58~\text{kHz},~4.3~\mu \text{s},~7~\text{~ms}^{-1})$---and predicts the last CPMG experiment with an order-of-magnitude smaller error, $\epsilon=0.02$.
We thus arrive at a minimally self-consistent model $S^{\NV}_{\text{min}}$, consistent with all observed qubit dynamics
\footnote{We remark that while $S_w$ was initially introduced as it is consistent with R-E dynamics, additional $S_{\CP}(\omega_m)$ independently reveal a nonvanishing baseline around $\min_m[S_{\CP}(\omega_m)]=5(1)\text{~ms}^{-1}$ [Fig.~\ref{FigSCNoiseModel}(a)]---in good agreement with $S_w$ predicted from an order of magnitude away with minimal experiment cost.}.

\begin{figure*}
\centering
\includegraphics[width=1.95\columnwidth]{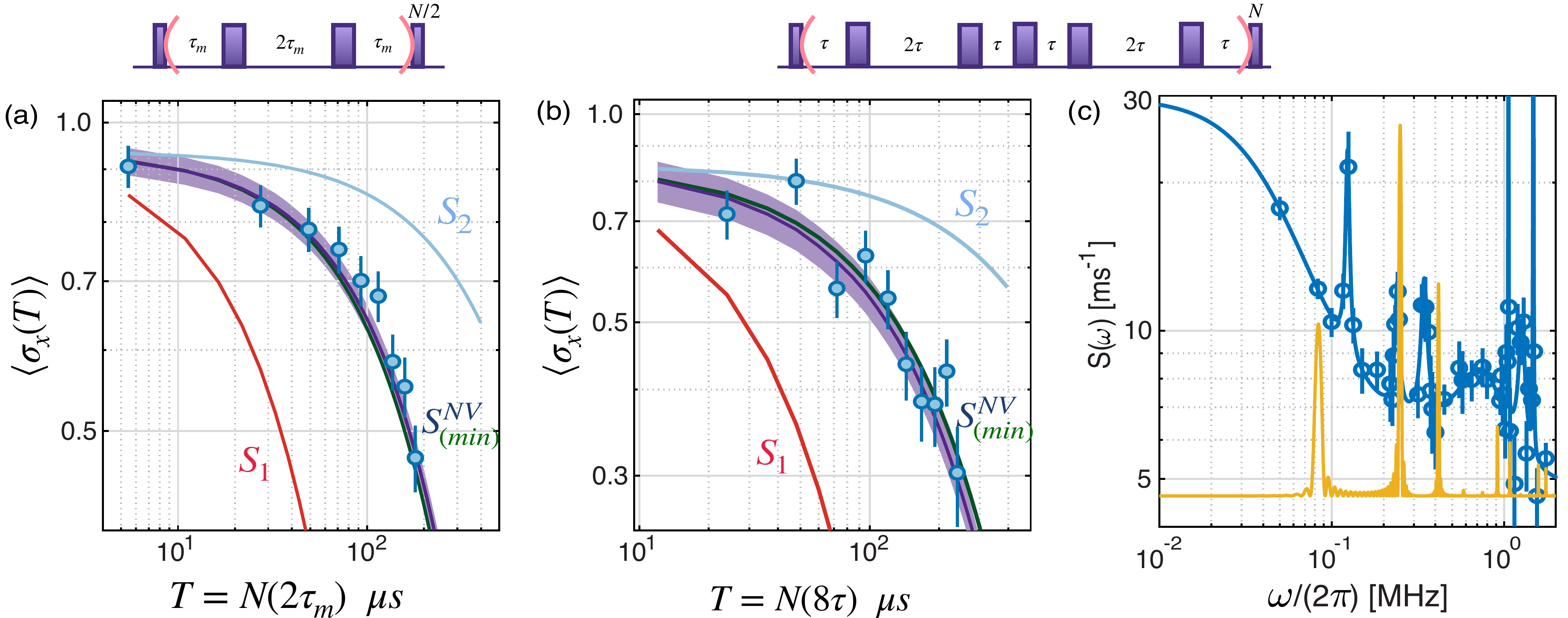}
\caption{
\textbf{Predictive power of the self-consistent noise model.}
{As a crucial check that the developed self-consistent noise model ${S}^{\NV}_{(\text{min})}(\omega)$ is an accurate model of the underlying bath}, we demonstrate that it is predictive even under new qubit dynamics, namely in (a) higher-frequency CPMG dynamics at $\omega_m\!=\!0.183~\text{MHz}$  (not used to inform ${S}^{\NV}_{\text{min}}$) as well as (b) Walsh dynamics of sequency $5$, while R-E-noise spectroscopy fails ($S_{1,2}$).
{(c) Walsh filter function.}
The Walsh filter $|\tilde{f}_{W(k,\lambda)}(\omega)|^2$ at time point $T=N\lambda=120~\mu \text{s}$ (gold) for sequency $k=5$ (cycle length $\lambda=8\tau$) is plotted against the noise spectrum directly measured from the data $S_{\CP}^{\NV}(\omega_m)$ and its fit (blue). The filter function  is scaled to more easily visualize which parts of the noise spectrum it is sampling. Here the filter was numerically generated by taking its finite-time Fourier transform of the Walsh time-domain function\cite{Stoffer91}.
}
\label{FigPredictivePower}
\end{figure*}

Additional $S_{\CP}(\omega_m)$ measurements can be used to further improve the model accuracy, either by revealing sharp resonances in the spectrum or by probing higher-frequency noise.
For our NV, $S_{\CP}(\omega_m)$ at higher-$\omega$ reveals multiple resolved peaks [Fig.~\ref{FigSCNoiseModel}(a)].
We thus obtain a final noise model ${S}^{\NV}$ by adding a series of spectral contributions $S_{pk}$ at $\omega_l$,
\begin{equation}\label{eq:SNV}
	{S}^{\NV}(\omega)  = \sum_{k=s,f} S_k(\omega|b_k,\tau_k)+ S_w + \sum_l S_{pk}(\omega-\omega_l).
\end{equation}
We remark that the same ${S}^{\NV}$  can be reached starting from CPMG experiments and achieving consistency with R-E decays.
Specifically, fitting the measured $S_{\CP}(\omega_m)$ yields Eq.~\ref{eq:SNV} \textit{minus} the slow-OU component ${S}_{s}$---since ${S}_{s}$ is narrow around $\omega=0$, it is only observed under R-E dynamics, while it is canceled out by CPMG pulse sequences.

{To summarize, having measured the qubit decoherence under various dynamics, we were able to define a self-consistent classical noise spectrum ${S}_{(\text{min})}^{\NV}(\omega)$,  which can self-consistently predict all of the already observed decoherence, as verified numerically (Fig.~\ref{FigSCNoiseModel}).
Now, as a crucial check that this noise spectrum  is  an accurate model of the underlying quantum environment, we also verify that it is predictive  of new qubit dynamics.
We first verify that ${S}_{\text{min}}^{\NV}(\omega)$  can predict new CPMG dynamics probing  order-of-magnitude higher frequencies [Fig.~\ref{FigPredictivePower}(a)]. Then, to probe a unique qubit dynamics, we perform a Walsh dynamical decoupling sequence of sequency 5 with asymmetric qubit-bath evolution times~\cite{Hayes11,Cooper14}, distinct from Ramsey, echo or CPMG sequences. Despite the more complicated dynamics [Fig.~\ref{FigPredictivePower}(c)], we verify that ${S}_{(\text{min})}^{\NV}(\omega)$ is predictive [Fig.~\ref{FigPredictivePower}(b)].
}

\subsection{X electronic spin qubit}\label{sec:X}
Having successfully characterized the noise experienced by the NV electronic spin, we turn to examine the noise of a nearby electronic spin X.
{Characterized in earlier works~\cite{Cooper19,Cooper20,Sun20}, the X spin is an electron-nuclear spin defect (each of spin 1/2) that is optically dark (at least with respect to $532~\text{nm}$ NV illumination). It is located several nanometers away from the NV with coupling strength $d \approx 60~\text{kHz}$~\cite{Cooper20}.} To achieve unitary control of the dark electronic spin X ($S,I=1/2$), we apply a two-tone microwave drive resonant with each of its hyperfine transitions  ($A^\X_{zz}\approx 26.4~\text{MHz}$  at the given field orientation\cite{Cooper20}). The Hartmann-Hahn protocol is exploited to achieve initialization and readout via the NV center~\cite{Cooper20}.

\begin{figure*}
\centering
\includegraphics[width=1.95\columnwidth]{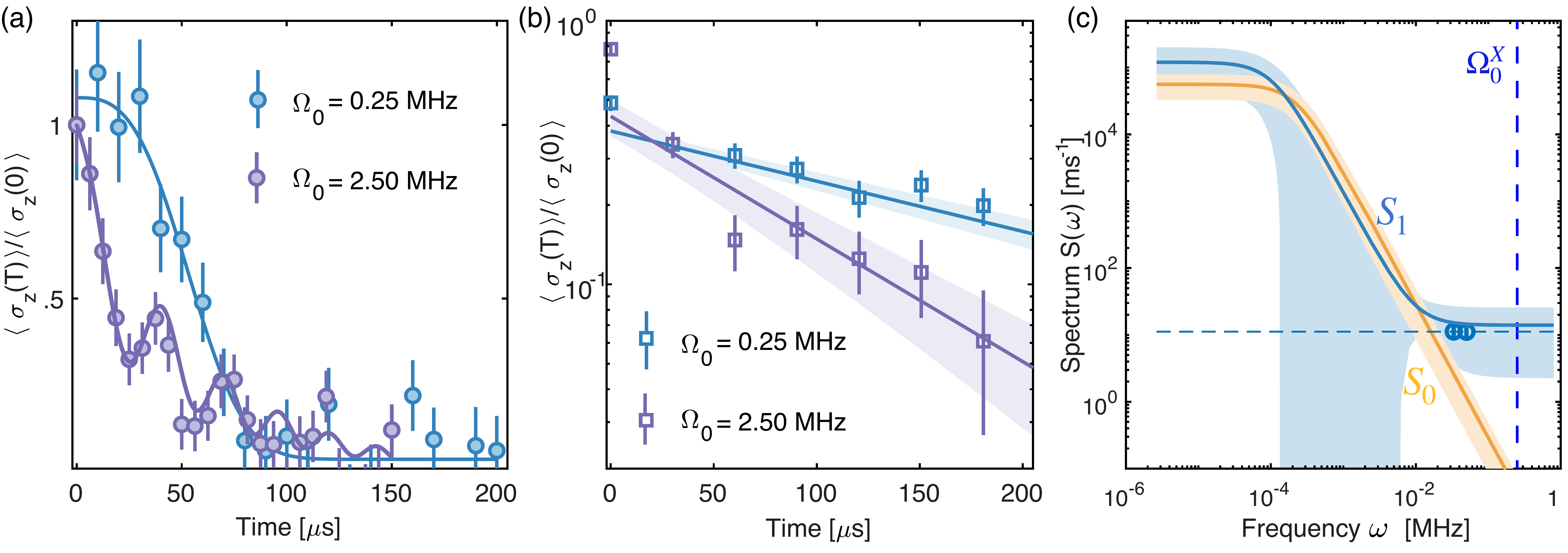}
\caption{
\textbf{Quantum bath of dark electronic spin X.}
{By investigating the spin echo (a) and CPMG (b) dynamics with varying driving (Rabi) strengths $\Omega$, we verify the quantum nature of the local environment of the X spin, realized by the presence of near-resonant and interacting (NRI) spins.
(a) X echo dynamics reveals the presence of NRI spins: At sufficiently high Rabi strength $\Omega_{h}$, small-amplitude oscillations are observed, akin to SEDOR and DEER experiments (the curve is a guide to the eye). However, at sufficiently low Rabi strength $\Omega_{l}$, the monotonic decay expected under single-qubit echo is recovered (the curve shows the fit to decays under ${S}^{\X}_{0,1}$ which overlap).
(b) X CPMG dynamics further reveals the effect of NRI spins, where typically one expects to observe higher coherence time $T_2$ with increasing Rabi power $\Omega$ due to higher control $\pi$ pulse fidelity $F_\pi(\Omega)$ (Appendix~\ref{appendix:Fpi}). However, here we observe the opposite behavior, where the higher $F_\pi$ leads to lower $T_2$, because the larger $\Omega_{h}$ recouples the X spin to a larger electronic-spin environment. (c) Finally, while the quantum bath precludes a classical noise model, by suppressing the X spin interaction with the quantum NRI spins–--which is possible by suppressing $\Omega_{}$–--we successfully recover a classical model for the X qubit over a restricted frequency range, following the same protocol as discussed with the NV.}
}
\label{FigXQuantumBath}
\end{figure*}

As the NV and X spins are in physical proximity of the same quantum environment, one may naively expect to find a self-consistent classical noise model for X, similar to that of the NV.
Instead, while we observe a monotonic Gaussian decay as expected under X Ramsey dynamics, small-amplitude oscillations appear under echo [Fig.~\ref{FigXQuantumBath}(a)] as well as multiple CPMG experiments. The presence of oscillations is inconsistent with either single-qubit dynamics or the exponential decay expected from an effectively classical bath, the prerequisite for a classical noise model.

To identify the cause of observed oscillations, we hypothesize the presence of near-resonant and interacting (NRI) spins around X (Fig.~\ref{FigScheme}). This behavior is indeed reminiscent of spin echo double resonance (SEDOR) experiments, where the control ($\pi$) pulses drive both spins to refocus their interaction, leading to signal oscillations at the frequency set by the interaction strength~\cite{deLange12,Belthangady13,Laraoui13,Sushkov14,Knowles16,Rosenfeld18,Cooper20,Pinto20,Degen21}.

To experimentally verify the presence of this complex spin environment, we study the echo dynamics of X at varying driving strengths.
Full-amplitude oscillations are not expected, since NRI spins are not driven on-resonance and do not experience a perfectly refocusing $\pi$ pulse.
Still, as the X Rabi-frequency $\Omega_0$ is increased beyond the detuning of the $k$th NRI spin from resonance, $|\Omega_0|>|\omega_k-\omega_0|$, we expect progressively effective driving and thus SEDOR oscillations. Conversely, at sufficiently weak Rabi frequency $|\Omega_0|\ll\min_k|\omega_k-\omega_0|$, as only the X qubit should be driven, we expect a monotonic decay.
To test this prediction, we measure the nominal X echo at two Rabi-frequencies, high $\Omega_{h} \!=\! 2.5~\text{MHz}$ and low $\Omega_{l}\! =\! \Omega_{h}/10$.
At $\Omega_{l}$ we observe monotonic decoherence, without oscillations, as expected of single-qubit dynamics in the presence of noise, while oscillations are only visible at $\Omega_{h}$ [Fig.~\ref{FigXQuantumBath}(a)].

Interestingly, the CPMG dynamics at varied driving strengths also reveals the effect of the NRI spins.
As suggested by prior experimental works~\cite{deLange12,Kucsko18,Degen21}, the presence of multiple NRI spins with different couplings can lead to faster decoherence when increasing $\Omega_0$  (since more spins become affected by the driving), effectively increasing the size of the spin environment by refocusing their interactions.
Performing X CPMG experiments, we indeed observe $T_2(\Omega_{h})\! <\! T_2(\Omega_{l})$ [Fig.~\ref{FigXQuantumBath}(b)]---despite $F_\pi(\Omega_{h})\!>\! F_\pi(\Omega_{l})$ (Appendix \ref{appendix:Fpi}). Crucially, in the absence of NRI spins we expect the opposite behavior, as stronger driving yields higher-fidelity $\pi$ pulses and can cancel the couplings to a broader range of noise sources.

Thus our experimental evidence strongly indicates the quantum nature of the environment of X.
Still, given the proximity of NV and X spins,
we expect both spins to interact with a largely similar environment, for which it was possible for the NV to develop an effective classical model
(indeed, the NV $\pi$ pulses are detuned by hundreds of megahertz due to its zero-field splitting).
We thus attempt to recover a classical noise model for X, by suppressing the quantum character of the spin environment by sufficiently reducing the Rabi power.
Using $\Omega_{l}$ to perform Ramsey and echo experiments, we obtain two minimal models,
${S}^{\X}_0(\omega)={S}_s(\omega|b_s^{\X},\tau_s^{\X})$, and ${S}^{\X}_1(\omega)={S}_s(\omega|b_s^{\X},\tau_s^{\X})+ {S}_w^{\X}$.
Following the same protocol as for the NV, we measure CPMG decays to verify which model is predictive.
Despite the severely restricted bandwidth, $S_{\CP}(\omega_m\ll \Omega_l)$, we are able to confirm the validity of ${S}^{\X}_1$, while ruling out ${S}^{\X}_0$ [Fig.~\ref{FigXQuantumBath}(c)].

\section{Discussion}\label{sec:discussion}
{Our results  point to a protocol for quantum sensing of complex many-body environments with nanoscale spatial resolution, achieved by comparing the common noise sources shared by nearby $n\geq2$ qubits. As a proof-of-principle demonstration,}
here we compare the dominant noise acting on both qubits, $S_s(\omega|b,\tau_c)$, arising from the quasistatic many-body electronic spin bath.
This reveals local bath properties with nanoscale spatial-resolution, not attainable by conventional NMR or an ensemble of single-qubit sensors.
First, the characteristic qubit-bath interaction strength $b$ reveals an estimate of the local spin density {(Appendix \ref{appendix:density})}, from which we estimate $f^{\NV}\!\approx\! 0.69(2)~\text{ppm}, ~f^{\X}\!\approx\! 0.22(2)~\text{ppm}$ from $S_s^{\NV}$ and $S_s^{\X}$, respectively.
Not only is this within the order of magnitude of the expected defect density given sample implantation ({Appendix \ref{appendix:spinbath}}),
but also importantly, the accurate estimate of $b$ reveals significant variation in the local spin density, even across nanometer lengthscales.
Similarly, the characteristic timescale of the noise process $\tau_c$ probes \textit{locally} the (qubit-independent) bath correlation time, determined by its internal evolution~\cite{Klauder62,Dobrovitski09,deLange12}.
For two qubits interacting with the same bath, we expect $\tau_c^{\NV}\!=\!\tau_c^{\X}$.
Interestingly, we observe instead a significant discrepancy, $\frac{\tau^{\NV}}{\tau^{\X}}\!=\!6(4)$, revealing that the spin bath properties at the nanoscale can vary significantly. Interestingly, this also contradicts a naive assumption of a bath of homogeneous spin species, for which we expect  ${(b\tau_c)^{\NV}}\!\approx \!{(b\tau_c)^{\X}}$, even accounting for varying spatial density as naively both $b \propto\! f$ and $\tau_c^{-1} \propto\! f$~\cite{Wang13,Kucsko18,Bauch20}.
Going further, we can attempt to explain the origin of the significant variations in $(b,\tau_c)$ at different spatial positions by a simple model.
{The observed stronger qubit-bath coupling $b$ for the NV, but with \textit{slower} bath fluctuation $\tau_c$ than for X, suggests the presence of a denser bath around the NV, but with considerable disorder (e.g., due to inhomogeneous spin species), which hinders energy-conserving spin flip-flop. Conversely, despite the lower density around X, there exist spins sufficiently nearer in resonance to result in faster flip-flops. This is in agreement with our discovery of NRI spins around X.}
{Thus one can envision that, given a spatial network of qubits at locations $\vec{x}_j$ (or a qubit on an atomic force microscope (AFM) tip \cite{Grinolds13,Myers14}), by measuring $(b,\tau_c)$ as a function of $\vec x$ it becomes possible to map out an unknown complex many-body spin environment, which reveals not only (quantitatively) the local spin density and effective decoherence time of the local spin bath but also (qualitatively) whether locally it is composed of a homogeneous spin species with either uniform or spatially varying density.}

\section{Conclusion and Outlook}\label{sec:outlook}

In this paper, we demonstrate a protocol to build a noise model that is not only self-consistent but also even predictive of qubit dynamics under varying controls, by reconciling complementary approaches to quantum noise spectroscopy.
Crucially, our method is strictly more accurate and robust compared with existing techniques, as it succeeds  even when other methods fail to yield the correct noise model.
Thanks to its simplicity and the potential to develop a practical yet predictive noise model of quantum devices,  our method can find application in various qubit platforms, further revealing interesting physical insights peculiar to each platform.

Extensions to multiqubit devices enables applications not only in quantum sensing but also in quantum device characterization. Indeed,
of significant interest is the characterization of correlated noise between qubits, which has implications for not only development of high-fidelity multiqubit (entangling) gates, but also practical realizability of quantum error-correction protocols~\cite{Szankowski16,Paz-Silva17,vonLupke20,Layden20}.
Our work contributes to the characterization of correlated noise, not only as common noise between qubits contributes to correlated noise, but also more importantly as accurate knowledge of individual-qubit noise is a prerequisite to reveal correlations~\cite{Szankowski16}.
As already demonstrated in this paper, the accurate characterization of noise at the single-qubit level can reveal a markedly non-uniform noise profile across a multiqubit processor (surprisingly, even across nanoscale distances), of which certain novel quantum protocols such as quantum error-corrected sensing schemes~\cite{Layden18b} can take advantage.

As a final remark, the absence of a self-consistent classical model heralds that the underlying bath is sufficiently complex, either of quantum or of non-Gaussian nature. In our system, we discover a quantum (possibly coherent) group of near-resonant electronic spins interacting with the X spin. Motivated by recent pioneering work in engineering larger quantum registers of electronic spins~\cite{Degen21}, we note that the system as observed here opens the door to building and controlling even larger electronic-spin registers—--beyond the coherence of the central qubit.

\acknowledgments
The authors thank Alexandre Cooper, Lorenza Viola, Guoqing Wang, Dirk Englund, William D. Oliver, Mingda Li, and Bilge Yildiz for invaluable and illuminating discussions. This work was in part supported by ARO Grant No. W911NF-11-1-0400 and by HRI-US.

\appendix

\section{Characterization of $\pi$-pulse Fidelity $F_\pi$}\label{appendix:Fpi}

\begin{figure}[t!]
\centering
\includegraphics[width=0.9\columnwidth]{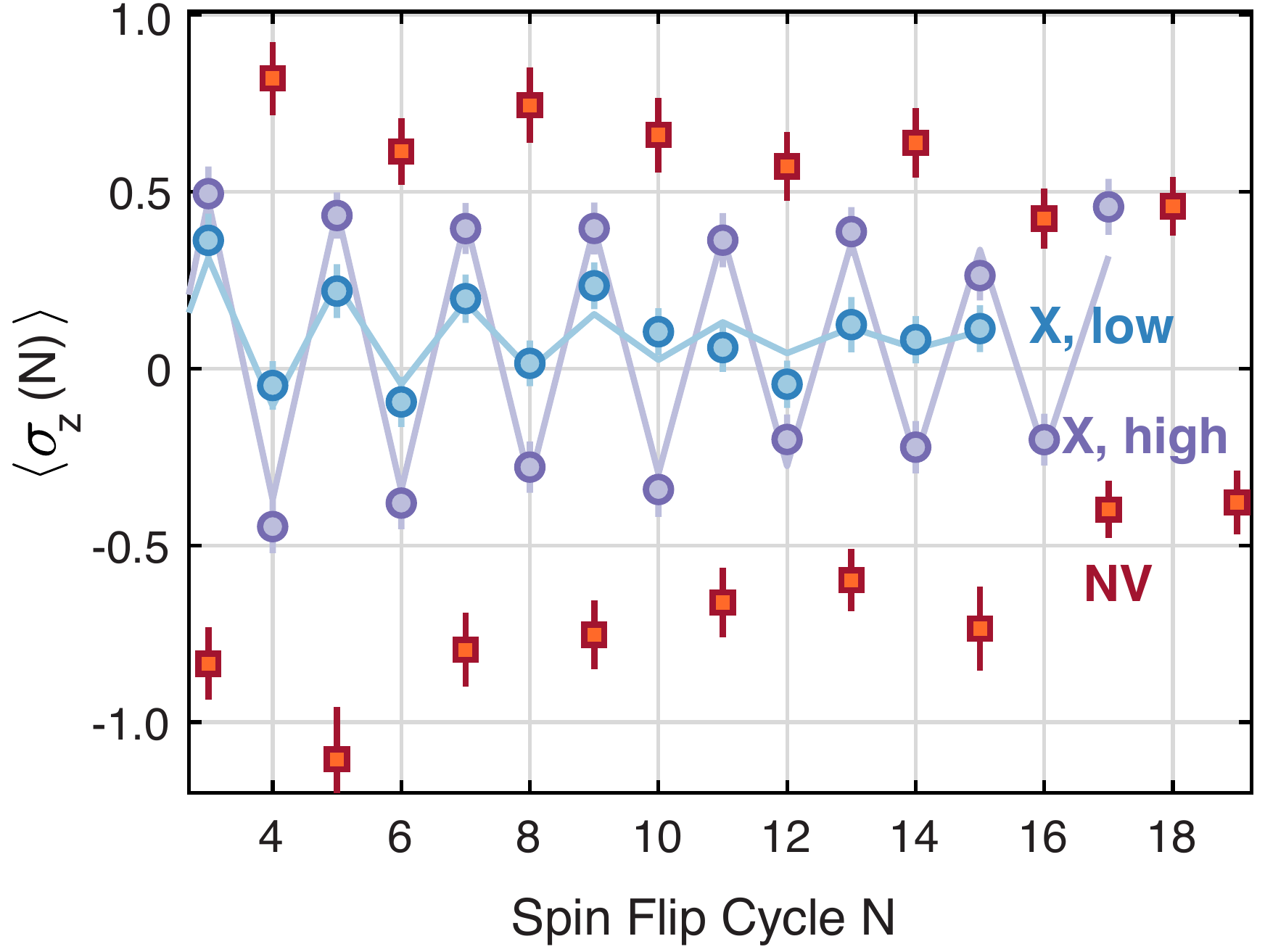}
\caption{
\textbf{Characterization of $\pi$-pulse fidelities $F_\pi$.}
The measured NV (red square) and X (circles) $\pi$-pulse fidelities $F_\pi$ are shown.
The NV is controlled by a single-tone $\pi$ pulse resonant with the $m_s=\{0,-1\}$ transition and strength $\Omega_0=6.76~\text{MHz}$. By fitting the signal to  $\langle\sigma_z(N)\rangle =  \beta_0 (-n_z)^N\!+\!c_0$, we extract $F_\pi\!=\!|\textrm{Tr}[U_\pi^\dag R]|/2\!= \!\sqrt{(1-n_z)/2}\!=0.987(0)$ for the NV. On the other hand, the X spin is modulated by a two-tone driving on resonance with the nuclear hyperfine splitting, to effectively remove the nuclear spin degree of freedom~\cite{Sun20}.
As expected, the control fidelity of the X spin is higher at higher Rabi frequency $\Omega_h\!=\!10\Omega_l \!=\! 2.5~\text{MHz}$: $F_\pi(\Omega_h)\!=\!0.992(2)$ (purple) $>F_\pi(\Omega_l)\!=\!0.955(7)$ (blue).
}
\label{FigPiPulseFidelity}
\end{figure}

In the main text we showed that the decay rate of the X spin under dynamical decoupling increased with the driving strength. As we expect that the decay rate should have contributions from the qubit-bath interactions during the free evolution and from the pulse imperfections,
\begin{align*}
	 T_{2}^{-1} & = T_{2,b}^{-1} + \gamma_c(F_\pi),
\end{align*}
we need to evaluate the driving fidelity (here the $\pi$-pulse fidelity $F_\pi$) in order to find  $T_{2,b}$, which characterizes the noise due to the bath alone. In general, for imperfect control, $F_\pi<1$, there is an additional decay due to imperfect pulses, which is detrimental when performing noise spectroscopy, since it might mask the correct shape of the noise spectrum.

Here, we use a simple method to experimentally characterize the $\pi$-pulse fidelity $F_\pi\!=|\textrm{Tr}[U_\pi^\dag R]|/2$, which is particularly useful in the presence of strong noise ($1/T_2^*\!\gg\!1$).
Here $U_\pi$ is the ideal $\pi$-pulse unitary and $R$ is the experimental one.
For a single qubit, an imperfect $\pi$-pulse rotation of duration $L$ might be due to a miscalibrated or fluctuating driving amplitude $\Omega_0$ or to an offset from resonance, $\delta$. The actual evolution is then
 $R \!=\! e^{-i (\Omega_0 \sigma_x + \delta \sigma_z)L/2}$, yielding $F_\pi = \langle\left | \left ( \frac{\Omega_0}{\Omega} \right ) \sin(\frac{\Omega L}{2}) \right |\rangle$, where $\Omega = \sqrt{(\Omega_0+\delta)}$.
 If the main pulse error arises from an off-resonance Hamiltonian, a larger driving strength will lead to better fidelity. However, if there are imperfections in the Rabi driving, typically larger driving results in larger deviations, and thus lower fidelities.

The experimental sequence we use to estimate $F_\pi$ is simply a series of $N$ spin flips, realized by (imperfect) $\pi$-pulses, applied to an initial population state $\rho_0=\frac12 (\mathbf{1}+ \beta_0 \sigma_z) $.
Importantly, each pulse is separated by interpulse delay $\tau \gg T_2^*$,
in order to ensure that any qubit coherence has decayed before the next $\pi$ pulse is applied, i.e., $\langle \sigma_{x(y)} \rangle = \Tr{\rho \sigma_{x(y)}} \rightarrow 0$, while the polarization $\langle \sigma_{z} \rangle$ should be ideally maintained.
In other words, at each cycle of unitary $\pi$ rotation $R$ and coherence decay, the state evolves as:
\begin{align*}
	 \rho & =R \rho_0 R^\dag = \frac12 (\openone + \beta_0 R\sigma_z R^\dag) \\
	 & = \frac12 (\openone + \beta_0 \vec n. \vec \sigma),\qquad ||\vec n||=1 \\
	 & \xrightarrow[]{\tau>T_2^*}  \frac12 (\openone + \beta_0 n_z \sigma_z).
\end{align*}
Then, the expectation value $\langle \sigma_z\rangle=\beta_0 n_z$ yields the fidelity, $n_z = \Tr{\sigma_zR\sigma_zR^\dag} = 1-2F_\pi^2$. For a more precise estimate, we vary the number of $\pi$ pulses, so that  after $N$ cycles the z-measurement yields $\langle \sigma_{z}(N) \rangle \approx \beta_0 n_z^N$.
Then, fitting the experimental data $\langle \sigma_{z}(N) \rangle$ to $\beta_0 (-n_z)^N$, one can directly estimate $F_\pi = \sqrt{(1-n_z)/2}$. Empirically, because $\tau\approx 2T_2^*$ suffices to ensure full decay of coherences, the method is useful for qubits under strong noise environments.

Figure ~\ref{FigPiPulseFidelity} shows that the control fidelity is better for higher driving strengths, as expected when off-resonant effects (including from noisy fields) are the main source of error. Then, we would also expect that higher fidelity pulses would also lead to slower decay.
Surprisingly, we find however that the overall measured decay time $T_{2}^{-1}(\Omega_h)>T_{2}^{-1}(\Omega_l)$ is shorter for higher-power driving, even if $F_\pi^{\X}(\Omega_h)>F_\pi^{\X}(\Omega_l)$ (Fig.~\ref{FigPiPulseFidelity}). This again indicates that  the noise from the bath $T_{2,b}^{-1}$ depends on the choice of driving power $\Omega_0^\X$, and in fact increases with $\Omega_0^\X$,
 while we can exclude the scenario where the higher driving power results in a reduced fidelity of the control pulses.
This observation is in agreement with our discovery of the NRI spins from the spin echo dynamics in the main text, consistent with the known SEDOR effect,  whereby either increasing the qubit driving power or selectively recoupling additional resonant spin groups resulted in stronger decoherence of the central qubit \cite{deLange12,Kucsko18,Degen21}.

\section{Physical origin of dominant noise for electronic spins}\label{appendix:spinbath}
The characteristics of the experimental system used in this paper (already introduced~\cite{Cooper19,Cooper20,Sun20})  provide insights into the physical origin of the observed noise.
The NV center was created via implantation of 14keV $^{15}$N ions with a dose of $10^{13}~\text{ions/cm}^2$ through a poly(methyl methacrylate) (PMMA) mask with 30nm diameter apertures deposited on top of a SiO$_2$ mask (to mitigate channeling effects) on an isotopically purified $^{12}$C diamond layer\cite{Cooper20}.
The relatively high implantation dose is expected to yield a high nitrogen concentration [N] and, due to limited N-to-NV conversion efficiency ($\sim5\%$ under annealing at 800K), only a few NVs per implantation spot. We note that of the $>\!150$ spots surveyed, only three (including the one investigated) had only one single NV, indicating potentially a smaller [N] or conversion efficiency. The implantation energy is expected to give an average  depth of $\sim\!20\pm7$~nm~\cite{Cooper20} [based on Stopping and Range of Ions in Matter (SRIM calculations], thus reducing, but potentially not eliminating, surface effects.
Therefore NV decoherence is expected to be limited by the  electronic spin bath \cite{Wang13} formed predominantly by N-related spin defects, with possible additional defects introduced from the mask or from the surface.
This is consistent with our  observation that the dominant noise experienced by the NV is given by a slow-OU noise $S_s$, characteristic of a quasistatic many-body electronic spin bath observed in Refs.~\cite{Dobrovitski08, deLange10, Wang13}.
The spin bath observed in our sample is however more complex than in these previous works, and our ability to probe it with two distinct spin probes a few nanometers apart provides additional insight into the bath properties and dynamics.

First, the double-OU noise $S_s+S_f$ observed for the NV suggests that there exist two distinct groups of electronic spin baths, distinguished by the timescale of their intrabath dynamics $\tau_{c}$.
A similar double-OU model has been used in an earlier work \cite{Myers14} to successfully describe the noise experienced by shallow NV centers in diamond, with $S_f$ attributed to the faster fluctuating spins on the surface. A similar scenario could describe our diamond, with the deeper NV resulting in the smaller $b_f\ll b_s$, while the observed $(b^f,\tau_c^f)$ agree within an order of magnitude of those reported in Ref. \cite{Myers14}.
In our sample, the NV still interacts more strongly with the bulk spin bath that we can probe now with nanoscale spatial resolution --- using another controllable electronic spin (X) several nanometers away from the NV.

Naively, due to the spatial proximity of NV and X spins, one may expect a largely similar noise experienced by both electronic spins. Surprisingly, we discover a local quantum environment around X, which precludes a classical description, realized by a group of near-resonant and interacting (NRI) electronic spins.
Still, by sufficiently suppressing the interaction between X and NRI spins, we uncover the  underlying dominant slow spin bath $S_s^\X$, as reported by the NV. The results highlight strong variations of the spin environment at the nanoscale (see Sec.~\ref{sec:discussion} and Appendix~\ref{appendix:noise}), further confirming the need for multiqubit noise spectroscopy.

\section{Spin bath properties derived from the observed noise spectrum}
\label{appendix:noise}
In the main text, we modeled the noise spectrum with a sum of Ornstein-Uhlenbeck (OU) noise processes, each given by an autocorrelation $\langle B(T)B(0) \rangle = b^2 e^{-T/\tau_c}$, fully characterized by two parameters $(b,\tau_c)$. Here we want to show how such a model can be related to the physical characteristics of spin baths.

\subsection{Local spin density from noise strength}\label{appendix:density}
The first parameter $b^2=\langle B^2(0) \rangle$ describes the  noise strength.
In the case of dephasing of a central qubit via the magnetic dipole interaction $H_{\text{int}}=S_z \sum_k^N A_k I_z^k$ to other spins, $b$ can help  estimate the local spin density.
The dipolar coupling strength $A_k$ between the central and $k$th spin is  $A_k = \frac{\mu_0 \gamma_e \gamma_k \hbar}{4\pi r_{k}^{3}} (1-3\cos^2(\theta_k))$, with $\gamma_{e(k)}$ being the gyromagnetic ratio of the central ($k$th) spins, $r_k$ being the inter-spin distance between the central and $k$th spin, and  $\theta_k$ the polar angle between $\vec r_k$ and the external magnetic field (assumed to be aligned with the zero-field splitting of the NV). Here we assume for simplicity $\gamma_k=\gamma_e$.

By defining the noise $H_{\text{int}}=B S_z$ with $B =\sum_k^N A_k I_z^k$ and
assuming the bath to be at thermal equilibrium, $\rho_B = \openone/2^N$, we can  replace the bath spin-1/2 operators with random variables and define the effective spin-qubit Hamiltonian  $H_{\text{int}}=B S_z$  with the random variable $B=\sum_k^N A_k I_z^k$ characterized by  $b^2$,
\begin{align}\label{eq:b2}
	b^2 & = \langle B^2(0) \rangle \nonumber\\
	    & = \left \langle \sum_k A_k^2 \openone/4 + \sum_{k\neq l} A_k A_l I_z^k I_z^l \right \rangle \nonumber\\
			& = \Tr \left [\rho_B  \left (\sum_k A_k^2 \openone/4 + \sum_{k\neq l} A_k A_l I_z^k I_z^l \right) \right] \nonumber \\
			& = \sum_k^N \frac14A_k^2.
\end{align}
We remark that $b^2$ is the second moment
$M_2 =\overline{(\Delta \omega^2)}_{SI}$ of the dipolar broadening by unlike spins~\cite{Abragam61}.

In the limit of a diluted spin bath ($f\ll 1$), we can replace the sum with an integral,
\begin{align*}
    b^2&=\int \frac14A^2(\vec r) \rho(\vec{r})d^3\vec{r},
\end{align*}
where $A(\vec r)=\frac{\mu_0 \gamma_e^2 \hbar}{4\pi r^3}(1-3\cos^2\theta)$ and we introduced the spin density $\rho ~(\text{cm}^{-3})$ [or atomic fraction $f ~(\text{ppm})$)]. We can thus estimate $\rho$ from the experimentally measured decoherence rate $b$,
\begin{align}
    b^2&=\frac{\mu_0^2 \gamma_e^4 \hbar^2}{4 (4\pi)^2} \left ( \frac{16\pi}{15} \int_{r_{\text{min}}}^R\frac\rho{r^4}dr \right ) \approx \frac{4\pi \mu_0^2 \gamma_e^4 \hbar^2}{(4\pi)^2 15} \frac{\rho}{r^3_{\text{min}}}
\end{align}
for sufficiently large $R^3\gg r^3_{\text{min}}$.
Here, $r_{\text{min}}$ should not be taken as the lattice constant, but instead it represents the typical inter-spin distance in the  sparse distribution of spins in the host lattice. We can assume that the probability of finding $n$ spins in a volume of radius $r$ is given by a Poisson distribution of mean $4\pi r^3\rho$. Then, following Ref.~\cite{Hoch88}, $r_{\text{min}}$ can be taken as the distance at which the probability of finding no other spin is 1/2, i.e., $p(x\!=\!0)=e^{-4\pi \rho r_{\text{min}}^3/3}=1/2$, which yields $r_{\text{min}}\approx 0.55 \rho^{-1/3}$. We finally have
\begin{align}
    b^2=\frac{4\pi \mu_0^2 \gamma_e^4 \hbar^2}{(4\pi)^2 15} \frac{\rho^2}{0.55^3}\approx 1.69\times10^{10}f^2 (\textrm{rad/s})^2,
\end{align}
from which we can estimate $f$ ($f= \frac{\rho ~(\text{cm}^{-3})}{1.77*10^{17} ~(\text{cm}^{-3})}$~ppm) from the experimental knowledge of $b$.
Our estimate, of $b\approx 0.13\times10^6 f$~rad/s, compares favorably with previous numerical results~\cite{Wang13}, which found $b\approx 0.78\times10^6 f$~rad/s.

We remark that this estimation of the density from the dephasing time (yielding a linear relationship, $T_2^{*,-1}\propto f$) is limited to sparse density $f<0.01$ \cite{Wang13}, while for sufficiently dense systems ($f\!>\!0.1$) one expects $T_2^{*,-1}\propto \sqrt{f}$ \cite{Abragam61}. Indeed, in that case  one can approximate the sum in Eq.~\ref{eq:b2} as
\begin{align*}
 b^2 & = \sum_k^N A_k^2/4  = f \sum_k' A_k^2/4 \equiv f A^2_{\text{tot}},
\end{align*}
where the prime indicates the sum over all lattice sites. For a known lattice structure in either one dimension (1D), 2D, or 3D, it is possible to (numerically) calculate the convergent sum $A^2_{\text{tot}}=\sum_k' A_k^2/4$, leaving $f$ as the only unknown. Similarly, even for sparse systems, one can evaluate the integral over other geometries, such as a 2D layer of surface spins \cite{Myers14,Romach15}.

\subsection{Disorder strength in the local bath of two spins}\label{app:disorder}
The auto-correlation time $\tau_c$, also called the correlation or `memory' time of the bath, describes the characteristic timescale of the noise fluctuation and is thus expected to be independent of the spin qubit used to probe the environment.

Even for a generic (quantum) bath,
The knowledge of $\tau_c$ may be of practical interest for a generic (even quantum) bath, e.g., to establish its Markovian character (indicated by  $\tau_c\rightarrow0$), which allows  modeling the qubit open system dynamics via a Lindblad master equation;  or  to investigate sources of correlated noise in a multiqubit device, more probable for  long-correlated noise sources, which is more difficult to analyze and correct.

For a spin bath,  the autocorrelation function $\langle B(T)B(0)\rangle$ describes the properties of the field generated by the spin bath configuration,  $B(T)=e^{i H_B T} B(0) e^{-i H_B T}$. The correlation time $\tau_c$, then,  characterizes the timescale over which $B(T)$ loses memory of its initial state $B(0) = \sum_k A_k I_z^k$, due to evolution under its internal dipolar Hamiltonian $H_B$, which leads to, e.g., spin flip-flops within the bath \cite{Dobrovitski09}. The correlation function can be often written  as an exponential decay, $\langle B(T)B(0)\rangle=b^2e^{-T/\tau_c}$, with $\tau_c^{-1} \equiv \sum_{j\!>\!k} R_{jk}$ given by the total spin flip-flop rate between all $j,k$ spin pairs $R_{jk} \propto A_{jk} \frac{\Gamma_d}{\Gamma_d^2+\delta^2}$ \cite{Bauch20}. Then, the correlation time depends not only on the spin density, $A_{jk}\propto f$, but also on the distribution of the spin frequencies. Indeed, the flip-flop rate $R_{jk}$  is  suppressed by frequency differences $\delta$ between each spin pair. Whereas $\delta$ is small for a homogeneous spin species, different hyperfine interactions (with strength on the order of megahertz),  can severely suppress the flip-flop via dipolar coupling (approximately kilohertz). Even dipolar coupling to other electronic spin species or to nuclear spins can quench the bath fluctuations~\cite{Bar-Gill12,Bauch20}.


\bibliographystyle{apsrev4-2}
\nocite{*}
\bibliography{mainPRB_final}

\begin{thebibliography}{61}%
\makeatletter
\providecommand \@ifxundefined [1]{%
 \@ifx{#1\undefined}
}%
\providecommand \@ifnum [1]{%
 \ifnum #1\expandafter \@firstoftwo
 \else \expandafter \@secondoftwo
 \fi
}%
\providecommand \@ifx [1]{%
 \ifx #1\expandafter \@firstoftwo
 \else \expandafter \@secondoftwo
 \fi
}%
\providecommand \natexlab [1]{#1}%
\providecommand \enquote  [1]{``#1''}%
\providecommand \bibnamefont  [1]{#1}%
\providecommand \bibfnamefont [1]{#1}%
\providecommand \citenamefont [1]{#1}%
\providecommand \href@noop [0]{\@secondoftwo}%
\providecommand \href [0]{\begingroup \@sanitize@url \@href}%
\providecommand \@href[1]{\@@startlink{#1}\@@href}%
\providecommand \@@href[1]{\endgroup#1\@@endlink}%
\providecommand \@sanitize@url [0]{\catcode `\\12\catcode `\$12\catcode
  `\&12\catcode `\#12\catcode `\^12\catcode `\_12\catcode `\%12\relax}%
\providecommand \@@startlink[1]{}%
\providecommand \@@endlink[0]{}%
\providecommand \url  [0]{\begingroup\@sanitize@url \@url }%
\providecommand \@url [1]{\endgroup\@href {#1}{\urlprefix }}%
\providecommand \urlprefix  [0]{URL }%
\providecommand \Eprint [0]{\href }%
\providecommand \doibase [0]{https://doi.org/}%
\providecommand \selectlanguage [0]{\@gobble}%
\providecommand \bibinfo  [0]{\@secondoftwo}%
\providecommand \bibfield  [0]{\@secondoftwo}%
\providecommand \translation [1]{[#1]}%
\providecommand \BibitemOpen [0]{}%
\providecommand \bibitemStop [0]{}%
\providecommand \bibitemNoStop [0]{.\EOS\space}%
\providecommand \EOS [0]{\spacefactor3000\relax}%
\providecommand \BibitemShut  [1]{\csname bibitem#1\endcsname}%
\let\auto@bib@innerbib\@empty
\bibitem [{\citenamefont {Goldstein}\ \emph {et~al.}(2011)\citenamefont
  {Goldstein}, \citenamefont {Cappellaro}, \citenamefont {Maze}, \citenamefont
  {Hodges}, \citenamefont {Jiang}, \citenamefont {S{\o}rensen},\ and\
  \citenamefont {Lukin}}]{Goldstein11}%
  \BibitemOpen
  \bibfield  {author} {\bibinfo {author} {\bibfnamefont {G.}~\bibnamefont
  {Goldstein}}, \bibinfo {author} {\bibfnamefont {P.}~\bibnamefont
  {Cappellaro}}, \bibinfo {author} {\bibfnamefont {J.~R.}\ \bibnamefont
  {Maze}}, \bibinfo {author} {\bibfnamefont {J.~S.}\ \bibnamefont {Hodges}},
  \bibinfo {author} {\bibfnamefont {L.}~\bibnamefont {Jiang}}, \bibinfo
  {author} {\bibfnamefont {A.~S.}\ \bibnamefont {S{\o}rensen}},\ and\ \bibinfo
  {author} {\bibfnamefont {M.~D.}\ \bibnamefont {Lukin}},\ }\href
  {https://doi.org/10.1103/PhysRevLett.106.140502} {\bibfield  {journal}
  {\bibinfo  {journal} {Physical Review Letters}\ }\textbf {\bibinfo {volume}
  {106}},\ \bibinfo {pages} {140502} (\bibinfo {year} {2011})}\BibitemShut
  {NoStop}%
\bibitem [{\citenamefont {Taminiau}\ \emph {et~al.}(2012)\citenamefont
  {Taminiau}, \citenamefont {Wagenaar}, \citenamefont {{van der Sar}},
  \citenamefont {Jelezko}, \citenamefont {Dobrovitski},\ and\ \citenamefont
  {Hanson}}]{Taminiau12}%
  \BibitemOpen
  \bibfield  {author} {\bibinfo {author} {\bibfnamefont {T.~H.}\ \bibnamefont
  {Taminiau}}, \bibinfo {author} {\bibfnamefont {J.~J.~T.}\ \bibnamefont
  {Wagenaar}}, \bibinfo {author} {\bibfnamefont {T.}~\bibnamefont {{van der
  Sar}}}, \bibinfo {author} {\bibfnamefont {F.}~\bibnamefont {Jelezko}},
  \bibinfo {author} {\bibfnamefont {V.~V.}\ \bibnamefont {Dobrovitski}},\ and\
  \bibinfo {author} {\bibfnamefont {R.}~\bibnamefont {Hanson}},\ }\href
  {https://doi.org/10.1103/PhysRevLett.109.137602} {\bibfield  {journal}
  {\bibinfo  {journal} {Physical Review Letters}\ }\textbf {\bibinfo {volume}
  {109}},\ \bibinfo {pages} {137602} (\bibinfo {year} {2012})}\BibitemShut
  {NoStop}%
\bibitem [{\citenamefont {Abobeih}\ \emph {et~al.}(2019)\citenamefont
  {Abobeih}, \citenamefont {Randall}, \citenamefont {Bradley}, \citenamefont
  {Bartling}, \citenamefont {Bakker}, \citenamefont {Degen}, \citenamefont
  {Markham}, \citenamefont {Twitchen},\ and\ \citenamefont
  {Taminiau}}]{Abobeih19a}%
  \BibitemOpen
  \bibfield  {author} {\bibinfo {author} {\bibfnamefont {M.~H.}\ \bibnamefont
  {Abobeih}}, \bibinfo {author} {\bibfnamefont {J.}~\bibnamefont {Randall}},
  \bibinfo {author} {\bibfnamefont {C.~E.}\ \bibnamefont {Bradley}}, \bibinfo
  {author} {\bibfnamefont {H.~P.}\ \bibnamefont {Bartling}}, \bibinfo {author}
  {\bibfnamefont {M.~A.}\ \bibnamefont {Bakker}}, \bibinfo {author}
  {\bibfnamefont {M.~J.}\ \bibnamefont {Degen}}, \bibinfo {author}
  {\bibfnamefont {M.}~\bibnamefont {Markham}}, \bibinfo {author} {\bibfnamefont
  {D.~J.}\ \bibnamefont {Twitchen}},\ and\ \bibinfo {author} {\bibfnamefont
  {T.~H.}\ \bibnamefont {Taminiau}},\ }\href
  {https://doi.org/10.1038/s41586-019-1834-7} {\bibfield  {journal} {\bibinfo
  {journal} {Nature}\ }\textbf {\bibinfo {volume} {576}},\ \bibinfo {pages}
  {411} (\bibinfo {year} {2019})}\BibitemShut {NoStop}%
\bibitem [{\citenamefont {Bradley}\ \emph {et~al.}(2019)\citenamefont
  {Bradley}, \citenamefont {Randall}, \citenamefont {Abobeih}, \citenamefont
  {Berrevoets}, \citenamefont {Degen}, \citenamefont {Bakker}, \citenamefont
  {Markham}, \citenamefont {Twitchen},\ and\ \citenamefont
  {Taminiau}}]{Bradley19}%
  \BibitemOpen
  \bibfield  {author} {\bibinfo {author} {\bibfnamefont {C.~E.}\ \bibnamefont
  {Bradley}}, \bibinfo {author} {\bibfnamefont {J.}~\bibnamefont {Randall}},
  \bibinfo {author} {\bibfnamefont {M.~H.}\ \bibnamefont {Abobeih}}, \bibinfo
  {author} {\bibfnamefont {R.~C.}\ \bibnamefont {Berrevoets}}, \bibinfo
  {author} {\bibfnamefont {M.~J.}\ \bibnamefont {Degen}}, \bibinfo {author}
  {\bibfnamefont {M.~A.}\ \bibnamefont {Bakker}}, \bibinfo {author}
  {\bibfnamefont {M.}~\bibnamefont {Markham}}, \bibinfo {author} {\bibfnamefont
  {D.~J.}\ \bibnamefont {Twitchen}},\ and\ \bibinfo {author} {\bibfnamefont
  {T.~H.}\ \bibnamefont {Taminiau}},\ }\href
  {https://doi.org/10.1103/PhysRevX.9.031045} {\bibfield  {journal} {\bibinfo
  {journal} {Physical Review X}\ }\textbf {\bibinfo {volume} {9}},\ \bibinfo
  {pages} {031045} (\bibinfo {year} {2019})}\BibitemShut {NoStop}%
\bibitem [{\citenamefont {Liu}\ \emph {et~al.}(2019)\citenamefont {Liu},
  \citenamefont {Ajoy},\ and\ \citenamefont {Cappellaro}}]{Liu19}%
  \BibitemOpen
  \bibfield  {author} {\bibinfo {author} {\bibfnamefont {Y.-X.}\ \bibnamefont
  {Liu}}, \bibinfo {author} {\bibfnamefont {A.}~\bibnamefont {Ajoy}},\ and\
  \bibinfo {author} {\bibfnamefont {P.}~\bibnamefont {Cappellaro}},\ }\href
  {https://doi.org/10.1103/PhysRevLett.122.100501} {\bibfield  {journal}
  {\bibinfo  {journal} {Physical Review Letters}\ }\textbf {\bibinfo {volume}
  {122}},\ \bibinfo {pages} {100501} (\bibinfo {year} {2019})}\BibitemShut
  {NoStop}%
\bibitem [{\citenamefont {Cooper}\ \emph {et~al.}(2020)\citenamefont {Cooper},
  \citenamefont {Sun}, \citenamefont {Jaskula},\ and\ \citenamefont
  {Cappellaro}}]{Cooper20}%
  \BibitemOpen
  \bibfield  {author} {\bibinfo {author} {\bibfnamefont {A.}~\bibnamefont
  {Cooper}}, \bibinfo {author} {\bibfnamefont {W.~K.~C.}\ \bibnamefont {Sun}},
  \bibinfo {author} {\bibfnamefont {J.-C.}\ \bibnamefont {Jaskula}},\ and\
  \bibinfo {author} {\bibfnamefont {P.}~\bibnamefont {Cappellaro}},\ }\href
  {https://doi.org/10.1103/PhysRevLett.124.083602} {\bibfield  {journal}
  {\bibinfo  {journal} {Physical Review Letters}\ }\textbf {\bibinfo {volume}
  {124}},\ \bibinfo {pages} {083602} (\bibinfo {year} {2020})}\BibitemShut
  {NoStop}%
\bibitem [{\citenamefont {Jackson}\ \emph {et~al.}(2021)\citenamefont
  {Jackson}, \citenamefont {Gangloff}, \citenamefont {Bodey}, \citenamefont
  {Zaporski}, \citenamefont {Bachorz}, \citenamefont {Clarke}, \citenamefont
  {Hugues}, \citenamefont {Le~Gall},\ and\ \citenamefont
  {Atat{\"u}re}}]{Jackson21}%
  \BibitemOpen
  \bibfield  {author} {\bibinfo {author} {\bibfnamefont {D.~M.}\ \bibnamefont
  {Jackson}}, \bibinfo {author} {\bibfnamefont {D.~A.}\ \bibnamefont
  {Gangloff}}, \bibinfo {author} {\bibfnamefont {J.~H.}\ \bibnamefont {Bodey}},
  \bibinfo {author} {\bibfnamefont {L.}~\bibnamefont {Zaporski}}, \bibinfo
  {author} {\bibfnamefont {C.}~\bibnamefont {Bachorz}}, \bibinfo {author}
  {\bibfnamefont {E.}~\bibnamefont {Clarke}}, \bibinfo {author} {\bibfnamefont
  {M.}~\bibnamefont {Hugues}}, \bibinfo {author} {\bibfnamefont
  {C.}~\bibnamefont {Le~Gall}},\ and\ \bibinfo {author} {\bibfnamefont
  {M.}~\bibnamefont {Atat{\"u}re}},\ }\href
  {https://doi.org/10.1038/s41567-020-01161-4} {\bibfield  {journal} {\bibinfo
  {journal} {Nature Physics}\ }\textbf {\bibinfo {volume} {17}},\ \bibinfo
  {pages} {585} (\bibinfo {year} {2021})}\BibitemShut {NoStop}%
\bibitem [{\citenamefont {Ruskuc}\ \emph {et~al.}(2022)\citenamefont {Ruskuc},
  \citenamefont {Wu}, \citenamefont {Rochman}, \citenamefont {Choi},\ and\
  \citenamefont {Faraon}}]{Ruskuc22}%
  \BibitemOpen
  \bibfield  {author} {\bibinfo {author} {\bibfnamefont {A.}~\bibnamefont
  {Ruskuc}}, \bibinfo {author} {\bibfnamefont {C.-J.}\ \bibnamefont {Wu}},
  \bibinfo {author} {\bibfnamefont {J.}~\bibnamefont {Rochman}}, \bibinfo
  {author} {\bibfnamefont {J.}~\bibnamefont {Choi}},\ and\ \bibinfo {author}
  {\bibfnamefont {A.}~\bibnamefont {Faraon}},\ }\href
  {https://doi.org/10.1038/s41586-021-04293-6} {\bibfield  {journal} {\bibinfo
  {journal} {Nature}\ }\textbf {\bibinfo {volume} {602}},\ \bibinfo {pages}
  {408} (\bibinfo {year} {2022})}\BibitemShut {NoStop}%
\bibitem [{\citenamefont {Crow}\ and\ \citenamefont {Joynt}(2014)}]{Crow14}%
  \BibitemOpen
  \bibfield  {author} {\bibinfo {author} {\bibfnamefont {D.}~\bibnamefont
  {Crow}}\ and\ \bibinfo {author} {\bibfnamefont {R.}~\bibnamefont {Joynt}},\
  }\href {https://doi.org/10.1103/PhysRevA.89.042123} {\bibfield  {journal}
  {\bibinfo  {journal} {Physical Review A}\ }\textbf {\bibinfo {volume} {89}},\
  \bibinfo {pages} {042123} (\bibinfo {year} {2014})}\BibitemShut {NoStop}%
\bibitem [{\citenamefont {Helm}\ \emph {et~al.}(2011)\citenamefont {Helm},
  \citenamefont {Strunz}, \citenamefont {Rietzler},\ and\ \citenamefont
  {W{\"u}rflinger}}]{Helm11}%
  \BibitemOpen
  \bibfield  {author} {\bibinfo {author} {\bibfnamefont {J.}~\bibnamefont
  {Helm}}, \bibinfo {author} {\bibfnamefont {W.~T.}\ \bibnamefont {Strunz}},
  \bibinfo {author} {\bibfnamefont {S.}~\bibnamefont {Rietzler}},\ and\
  \bibinfo {author} {\bibfnamefont {L.~E.}\ \bibnamefont {W{\"u}rflinger}},\
  }\href {https://doi.org/10.1103/PhysRevA.83.042103} {\bibfield  {journal}
  {\bibinfo  {journal} {Physical Review A}\ }\textbf {\bibinfo {volume} {83}},\
  \bibinfo {pages} {042103} (\bibinfo {year} {2011})}\BibitemShut {NoStop}%
\bibitem [{\citenamefont {Helm}\ and\ \citenamefont {Strunz}(2009)}]{Helm09}%
  \BibitemOpen
  \bibfield  {author} {\bibinfo {author} {\bibfnamefont {J.}~\bibnamefont
  {Helm}}\ and\ \bibinfo {author} {\bibfnamefont {W.~T.}\ \bibnamefont
  {Strunz}},\ }\href {https://doi.org/10.1103/PhysRevA.80.042108} {\bibfield
  {journal} {\bibinfo  {journal} {Physical Review A}\ }\textbf {\bibinfo
  {volume} {80}},\ \bibinfo {pages} {042108} (\bibinfo {year}
  {2009})}\BibitemShut {NoStop}%
\bibitem [{\citenamefont {Dobrovitski}\ \emph {et~al.}(2008)\citenamefont
  {Dobrovitski}, \citenamefont {Feiguin}, \citenamefont {Awschalom},\ and\
  \citenamefont {Hanson}}]{Dobrovitski08}%
  \BibitemOpen
  \bibfield  {author} {\bibinfo {author} {\bibfnamefont {V.~V.}\ \bibnamefont
  {Dobrovitski}}, \bibinfo {author} {\bibfnamefont {A.~E.}\ \bibnamefont
  {Feiguin}}, \bibinfo {author} {\bibfnamefont {D.~D.}\ \bibnamefont
  {Awschalom}},\ and\ \bibinfo {author} {\bibfnamefont {R.}~\bibnamefont
  {Hanson}},\ }\href {https://doi.org/10.1103/PhysRevB.77.245212} {\bibfield
  {journal} {\bibinfo  {journal} {Physical Review B}\ }\textbf {\bibinfo
  {volume} {77}},\ \bibinfo {pages} {245212} (\bibinfo {year}
  {2008})}\BibitemShut {NoStop}%
\bibitem [{\citenamefont {{de Lange}}\ \emph {et~al.}(2010)\citenamefont {{de
  Lange}}, \citenamefont {Wang}, \citenamefont {Rist{\`e}}, \citenamefont
  {Dobrovitski},\ and\ \citenamefont {Hanson}}]{deLange10}%
  \BibitemOpen
  \bibfield  {author} {\bibinfo {author} {\bibfnamefont {G.}~\bibnamefont {{de
  Lange}}}, \bibinfo {author} {\bibfnamefont {Z.~H.}\ \bibnamefont {Wang}},
  \bibinfo {author} {\bibfnamefont {D.}~\bibnamefont {Rist{\`e}}}, \bibinfo
  {author} {\bibfnamefont {V.~V.}\ \bibnamefont {Dobrovitski}},\ and\ \bibinfo
  {author} {\bibfnamefont {R.}~\bibnamefont {Hanson}},\ }\href
  {https://doi.org/10.1126/science.1192739} {\bibfield  {journal} {\bibinfo
  {journal} {Science}\ }\textbf {\bibinfo {volume} {330}},\ \bibinfo {pages}
  {60} (\bibinfo {year} {2010})}\BibitemShut {NoStop}%
\bibitem [{\citenamefont {{de Lange}}\ \emph {et~al.}(2012)\citenamefont {{de
  Lange}}, \citenamefont {{van der Sar}}, \citenamefont {Blok}, \citenamefont
  {Wang}, \citenamefont {Dobrovitski},\ and\ \citenamefont
  {Hanson}}]{deLange12}%
  \BibitemOpen
  \bibfield  {author} {\bibinfo {author} {\bibfnamefont {G.}~\bibnamefont {{de
  Lange}}}, \bibinfo {author} {\bibfnamefont {T.}~\bibnamefont {{van der
  Sar}}}, \bibinfo {author} {\bibfnamefont {M.}~\bibnamefont {Blok}}, \bibinfo
  {author} {\bibfnamefont {Z.-H.}\ \bibnamefont {Wang}}, \bibinfo {author}
  {\bibfnamefont {V.}~\bibnamefont {Dobrovitski}},\ and\ \bibinfo {author}
  {\bibfnamefont {R.}~\bibnamefont {Hanson}},\ }\href
  {https://doi.org/10.1038/srep00382} {\bibfield  {journal} {\bibinfo
  {journal} {Scientific Reports}\ }\textbf {\bibinfo {volume} {2}},\ \bibinfo
  {pages} {382} (\bibinfo {year} {2012})}\BibitemShut {NoStop}%
\bibitem [{\citenamefont {Yan}\ \emph {et~al.}(2013)\citenamefont {Yan},
  \citenamefont {Gustavsson}, \citenamefont {Bylander}, \citenamefont {Jin},
  \citenamefont {Yoshihara}, \citenamefont {Cory}, \citenamefont {Nakamura},
  \citenamefont {Orlando},\ and\ \citenamefont {Oliver}}]{Yan13}%
  \BibitemOpen
  \bibfield  {author} {\bibinfo {author} {\bibfnamefont {F.}~\bibnamefont
  {Yan}}, \bibinfo {author} {\bibfnamefont {S.}~\bibnamefont {Gustavsson}},
  \bibinfo {author} {\bibfnamefont {J.}~\bibnamefont {Bylander}}, \bibinfo
  {author} {\bibfnamefont {X.}~\bibnamefont {Jin}}, \bibinfo {author}
  {\bibfnamefont {F.}~\bibnamefont {Yoshihara}}, \bibinfo {author}
  {\bibfnamefont {D.~G.}\ \bibnamefont {Cory}}, \bibinfo {author}
  {\bibfnamefont {Y.}~\bibnamefont {Nakamura}}, \bibinfo {author}
  {\bibfnamefont {T.~P.}\ \bibnamefont {Orlando}},\ and\ \bibinfo {author}
  {\bibfnamefont {W.~D.}\ \bibnamefont {Oliver}},\ }\href
  {https://doi.org/10.1038/ncomms3337} {\bibfield  {journal} {\bibinfo
  {journal} {Nature Communications}\ }\textbf {\bibinfo {volume} {4}},\
  \bibinfo {pages} {2337} (\bibinfo {year} {2013})}\BibitemShut {NoStop}%
\bibitem [{\citenamefont {Wang}\ \emph {et~al.}(2020)\citenamefont {Wang},
  \citenamefont {Liu},\ and\ \citenamefont {Cappellaro}}]{Wang20}%
  \BibitemOpen
  \bibfield  {author} {\bibinfo {author} {\bibfnamefont {G.}~\bibnamefont
  {Wang}}, \bibinfo {author} {\bibfnamefont {Y.-X.}\ \bibnamefont {Liu}},\ and\
  \bibinfo {author} {\bibfnamefont {P.}~\bibnamefont {Cappellaro}},\ }\href
  {https://doi.org/10.1088/1367-2630/abd2e5} {\bibfield  {journal} {\bibinfo
  {journal} {New Journal of Physics}\ }\textbf {\bibinfo {volume} {22}},\
  \bibinfo {pages} {123045} (\bibinfo {year} {2020})}\BibitemShut {NoStop}%
\bibitem [{\citenamefont {Wang}\ \emph {et~al.}(2021)\citenamefont {Wang},
  \citenamefont {Liu}, \citenamefont {Zhu},\ and\ \citenamefont
  {Cappellaro}}]{Wang21}%
  \BibitemOpen
  \bibfield  {author} {\bibinfo {author} {\bibfnamefont {G.}~\bibnamefont
  {Wang}}, \bibinfo {author} {\bibfnamefont {Y.-X.}\ \bibnamefont {Liu}},
  \bibinfo {author} {\bibfnamefont {Y.}~\bibnamefont {Zhu}},\ and\ \bibinfo
  {author} {\bibfnamefont {P.}~\bibnamefont {Cappellaro}},\ }\href
  {https://doi.org/10.1021/acs.nanolett.1c01165} {\bibfield  {journal}
  {\bibinfo  {journal} {Nano Letters}\ }\textbf {\bibinfo {volume} {21}},\
  \bibinfo {pages} {5143} (\bibinfo {year} {2021})}\BibitemShut {NoStop}%
\bibitem [{\citenamefont {Cywi{\'n}ski}\ \emph {et~al.}(2008)\citenamefont
  {Cywi{\'n}ski}, \citenamefont {Lutchyn}, \citenamefont {Nave},\ and\
  \citenamefont {Das~Sarma}}]{Cywinski08}%
  \BibitemOpen
  \bibfield  {author} {\bibinfo {author} {\bibfnamefont {{\L}.}~\bibnamefont
  {Cywi{\'n}ski}}, \bibinfo {author} {\bibfnamefont {R.~M.}\ \bibnamefont
  {Lutchyn}}, \bibinfo {author} {\bibfnamefont {C.~P.}\ \bibnamefont {Nave}},\
  and\ \bibinfo {author} {\bibfnamefont {S.}~\bibnamefont {Das~Sarma}},\ }\href
  {https://doi.org/10.1103/PhysRevB.77.174509} {\bibfield  {journal} {\bibinfo
  {journal} {Physical Review B}\ }\textbf {\bibinfo {volume} {77}},\ \bibinfo
  {pages} {174509} (\bibinfo {year} {2008})}\BibitemShut {NoStop}%
\bibitem [{\citenamefont {{\'A}lvarez}\ and\ \citenamefont
  {Suter}(2011)}]{Alvarez11}%
  \BibitemOpen
  \bibfield  {author} {\bibinfo {author} {\bibfnamefont {G.~A.}\ \bibnamefont
  {{\'A}lvarez}}\ and\ \bibinfo {author} {\bibfnamefont {D.}~\bibnamefont
  {Suter}},\ }\href {https://doi.org/10.1103/PhysRevLett.107.230501} {\bibfield
   {journal} {\bibinfo  {journal} {Physical Review Letters}\ }\textbf {\bibinfo
  {volume} {107}},\ \bibinfo {pages} {230501} (\bibinfo {year}
  {2011})}\BibitemShut {NoStop}%
\bibitem [{\citenamefont {Yuge}\ \emph {et~al.}(2011)\citenamefont {Yuge},
  \citenamefont {Sasaki},\ and\ \citenamefont {Hirayama}}]{Yuge11}%
  \BibitemOpen
  \bibfield  {author} {\bibinfo {author} {\bibfnamefont {T.}~\bibnamefont
  {Yuge}}, \bibinfo {author} {\bibfnamefont {S.}~\bibnamefont {Sasaki}},\ and\
  \bibinfo {author} {\bibfnamefont {Y.}~\bibnamefont {Hirayama}},\ }\href
  {https://doi.org/10.1103/PhysRevLett.107.170504} {\bibfield  {journal}
  {\bibinfo  {journal} {Physical Review Letters}\ }\textbf {\bibinfo {volume}
  {107}},\ \bibinfo {pages} {170504} (\bibinfo {year} {2011})}\BibitemShut
  {NoStop}%
\bibitem [{\citenamefont {Ryan}\ \emph {et~al.}(2010)\citenamefont {Ryan},
  \citenamefont {Hodges},\ and\ \citenamefont {Cory}}]{Ryan10}%
  \BibitemOpen
  \bibfield  {author} {\bibinfo {author} {\bibfnamefont {C.~A.}\ \bibnamefont
  {Ryan}}, \bibinfo {author} {\bibfnamefont {J.~S.}\ \bibnamefont {Hodges}},\
  and\ \bibinfo {author} {\bibfnamefont {D.~G.}\ \bibnamefont {Cory}},\ }\href
  {https://doi.org/10.1103/PhysRevLett.105.200402} {\bibfield  {journal}
  {\bibinfo  {journal} {Physical Review Letters}\ }\textbf {\bibinfo {volume}
  {105}},\ \bibinfo {pages} {200402} (\bibinfo {year} {2010})}\BibitemShut
  {NoStop}%
\bibitem [{\citenamefont {Paladino}\ \emph {et~al.}(2002)\citenamefont
  {Paladino}, \citenamefont {Faoro}, \citenamefont {Falci},\ and\ \citenamefont
  {Fazio}}]{Paladino02}%
  \BibitemOpen
  \bibfield  {author} {\bibinfo {author} {\bibfnamefont {E.}~\bibnamefont
  {Paladino}}, \bibinfo {author} {\bibfnamefont {L.}~\bibnamefont {Faoro}},
  \bibinfo {author} {\bibfnamefont {G.}~\bibnamefont {Falci}},\ and\ \bibinfo
  {author} {\bibfnamefont {R.}~\bibnamefont {Fazio}},\ }\href
  {https://doi.org/10.1103/PhysRevLett.88.228304} {\bibfield  {journal}
  {\bibinfo  {journal} {Physical Review Letters}\ }\textbf {\bibinfo {volume}
  {88}},\ \bibinfo {pages} {228304} (\bibinfo {year} {2002})}\BibitemShut
  {NoStop}%
\bibitem [{\citenamefont {Galperin}\ \emph {et~al.}(2006)\citenamefont
  {Galperin}, \citenamefont {Altshuler}, \citenamefont {Bergli},\ and\
  \citenamefont {Shantsev}}]{Galperin06}%
  \BibitemOpen
  \bibfield  {author} {\bibinfo {author} {\bibfnamefont {Y.~M.}\ \bibnamefont
  {Galperin}}, \bibinfo {author} {\bibfnamefont {B.~L.}\ \bibnamefont
  {Altshuler}}, \bibinfo {author} {\bibfnamefont {J.}~\bibnamefont {Bergli}},\
  and\ \bibinfo {author} {\bibfnamefont {D.~V.}\ \bibnamefont {Shantsev}},\
  }\href {https://doi.org/10.1103/PhysRevLett.96.097009} {\bibfield  {journal}
  {\bibinfo  {journal} {Physical Review Letters}\ }\textbf {\bibinfo {volume}
  {96}},\ \bibinfo {pages} {097009} (\bibinfo {year} {2006})}\BibitemShut
  {NoStop}%
\bibitem [{\citenamefont {Bergli}\ and\ \citenamefont
  {Faoro}(2007)}]{Bergli07}%
  \BibitemOpen
  \bibfield  {author} {\bibinfo {author} {\bibfnamefont {J.}~\bibnamefont
  {Bergli}}\ and\ \bibinfo {author} {\bibfnamefont {L.}~\bibnamefont {Faoro}},\
  }\href {https://doi.org/10.1103/PhysRevB.75.054515} {\bibfield  {journal}
  {\bibinfo  {journal} {Physical Review B}\ }\textbf {\bibinfo {volume} {75}},\
  \bibinfo {pages} {054515} (\bibinfo {year} {2007})}\BibitemShut {NoStop}%
\bibitem [{\citenamefont {Chen}\ \emph {et~al.}(2018)\citenamefont {Chen},
  \citenamefont {Sun}, \citenamefont {Saha}, \citenamefont {Jaskula},\ and\
  \citenamefont {Cappellaro}}]{Chen18}%
  \BibitemOpen
  \bibfield  {author} {\bibinfo {author} {\bibfnamefont {M.}~\bibnamefont
  {Chen}}, \bibinfo {author} {\bibfnamefont {W.~K.~C.}\ \bibnamefont {Sun}},
  \bibinfo {author} {\bibfnamefont {K.}~\bibnamefont {Saha}}, \bibinfo {author}
  {\bibfnamefont {J.-C.}\ \bibnamefont {Jaskula}},\ and\ \bibinfo {author}
  {\bibfnamefont {P.}~\bibnamefont {Cappellaro}},\ }\href
  {https://doi.org/10.1088/1367-2630/aac542} {\bibfield  {journal} {\bibinfo
  {journal} {New Journal of Physics}\ }\textbf {\bibinfo {volume} {20}},\
  \bibinfo {pages} {063011} (\bibinfo {year} {2018})}\BibitemShut {NoStop}%
\bibitem [{\citenamefont {Wang}\ and\ \citenamefont
  {Takahashi}(2013)}]{Wang13}%
  \BibitemOpen
  \bibfield  {author} {\bibinfo {author} {\bibfnamefont {Z.-H.}\ \bibnamefont
  {Wang}}\ and\ \bibinfo {author} {\bibfnamefont {S.}~\bibnamefont
  {Takahashi}},\ }\href {https://doi.org/10.1103/PhysRevB.87.115122} {\bibfield
   {journal} {\bibinfo  {journal} {Physical Review B}\ }\textbf {\bibinfo
  {volume} {87}},\ \bibinfo {pages} {115122} (\bibinfo {year}
  {2013})}\BibitemShut {NoStop}%
\bibitem [{\citenamefont {Sza{\'n}kowski}\ and\ \citenamefont
  {Cywi{\'n}ski}(2018)}]{Szankowski18}%
  \BibitemOpen
  \bibfield  {author} {\bibinfo {author} {\bibfnamefont {P.}~\bibnamefont
  {Sza{\'n}kowski}}\ and\ \bibinfo {author} {\bibfnamefont {{\L}.}~\bibnamefont
  {Cywi{\'n}ski}},\ }\href {https://doi.org/10.1103/PhysRevA.97.032101}
  {\bibfield  {journal} {\bibinfo  {journal} {Physical Review A}\ }\textbf
  {\bibinfo {volume} {97}},\ \bibinfo {pages} {032101} (\bibinfo {year}
  {2018})}\BibitemShut {NoStop}%
\bibitem [{\citenamefont {{Hern{\'a}ndez-G{\'o}mez}}\ \emph
  {et~al.}(2018)\citenamefont {{Hern{\'a}ndez-G{\'o}mez}}, \citenamefont
  {Poggiali}, \citenamefont {Cappellaro},\ and\ \citenamefont
  {Fabbri}}]{Hernandez-Gomez18}%
  \BibitemOpen
  \bibfield  {author} {\bibinfo {author} {\bibfnamefont {S.}~\bibnamefont
  {{Hern{\'a}ndez-G{\'o}mez}}}, \bibinfo {author} {\bibfnamefont
  {F.}~\bibnamefont {Poggiali}}, \bibinfo {author} {\bibfnamefont
  {P.}~\bibnamefont {Cappellaro}},\ and\ \bibinfo {author} {\bibfnamefont
  {N.}~\bibnamefont {Fabbri}},\ }\href
  {https://doi.org/10.1103/PhysRevB.98.214307} {\bibfield  {journal} {\bibinfo
  {journal} {Physical Review B}\ }\textbf {\bibinfo {volume} {98}},\ \bibinfo
  {pages} {214307} (\bibinfo {year} {2018})}\BibitemShut {NoStop}%
\bibitem [{\citenamefont {Norris}\ \emph {et~al.}(2016)\citenamefont {Norris},
  \citenamefont {{Paz-Silva}},\ and\ \citenamefont {Viola}}]{Norris16}%
  \BibitemOpen
  \bibfield  {author} {\bibinfo {author} {\bibfnamefont {L.~M.}\ \bibnamefont
  {Norris}}, \bibinfo {author} {\bibfnamefont {G.~A.}\ \bibnamefont
  {{Paz-Silva}}},\ and\ \bibinfo {author} {\bibfnamefont {L.}~\bibnamefont
  {Viola}},\ }\href {https://doi.org/10.1103/PhysRevLett.116.150503} {\bibfield
   {journal} {\bibinfo  {journal} {Physical Review Letters}\ }\textbf {\bibinfo
  {volume} {116}},\ \bibinfo {pages} {150503} (\bibinfo {year}
  {2016})}\BibitemShut {NoStop}%
\bibitem [{\citenamefont {Kwiatkowski}\ and\ \citenamefont
  {Cywi{\'n}ski}(2018)}]{Kwiatkowski18}%
  \BibitemOpen
  \bibfield  {author} {\bibinfo {author} {\bibfnamefont {D.}~\bibnamefont
  {Kwiatkowski}}\ and\ \bibinfo {author} {\bibfnamefont {{\L}.}~\bibnamefont
  {Cywi{\'n}ski}},\ }\href {https://doi.org/10.1103/PhysRevB.98.155202}
  {\bibfield  {journal} {\bibinfo  {journal} {Physical Review B}\ }\textbf
  {\bibinfo {volume} {98}},\ \bibinfo {pages} {155202} (\bibinfo {year}
  {2018})}\BibitemShut {NoStop}%
\bibitem [{\citenamefont {Sung}\ \emph {et~al.}(2019)\citenamefont {Sung},
  \citenamefont {Beaudoin}, \citenamefont {Norris}, \citenamefont {Yan},
  \citenamefont {Kim}, \citenamefont {Qiu}, \citenamefont {{von L{\"u}pke}},
  \citenamefont {Yoder}, \citenamefont {Orlando}, \citenamefont {Gustavsson},
  \citenamefont {Viola},\ and\ \citenamefont {Oliver}}]{Sung19}%
  \BibitemOpen
  \bibfield  {author} {\bibinfo {author} {\bibfnamefont {Y.}~\bibnamefont
  {Sung}}, \bibinfo {author} {\bibfnamefont {F.}~\bibnamefont {Beaudoin}},
  \bibinfo {author} {\bibfnamefont {L.~M.}\ \bibnamefont {Norris}}, \bibinfo
  {author} {\bibfnamefont {F.}~\bibnamefont {Yan}}, \bibinfo {author}
  {\bibfnamefont {D.~K.}\ \bibnamefont {Kim}}, \bibinfo {author} {\bibfnamefont
  {J.~Y.}\ \bibnamefont {Qiu}}, \bibinfo {author} {\bibfnamefont
  {U.}~\bibnamefont {{von L{\"u}pke}}}, \bibinfo {author} {\bibfnamefont
  {J.~L.}\ \bibnamefont {Yoder}}, \bibinfo {author} {\bibfnamefont {T.~P.}\
  \bibnamefont {Orlando}}, \bibinfo {author} {\bibfnamefont {S.}~\bibnamefont
  {Gustavsson}}, \bibinfo {author} {\bibfnamefont {L.}~\bibnamefont {Viola}},\
  and\ \bibinfo {author} {\bibfnamefont {W.~D.}\ \bibnamefont {Oliver}},\
  }\href {https://doi.org/10.1038/s41467-019-11699-4} {\bibfield  {journal}
  {\bibinfo  {journal} {Nature Communications}\ }\textbf {\bibinfo {volume}
  {10}},\ \bibinfo {pages} {3715} (\bibinfo {year} {2019})}\BibitemShut
  {NoStop}%
\bibitem [{SOM()}]{SOM}%
  \BibitemOpen
  \href@noop {} {}\bibinfo {note} {See Supplemental Online Material
  at}\BibitemShut {NoStop}%
\bibitem [{Note1()}]{Note1}%
  \BibitemOpen
  \bibinfo {note} {We remark that a simpler noise model candidate $S_0\protect
  \! \propto \protect \!\delta (\omega )$, which also yields a gaussian Ramsey
  decay, has been ruled out as inconsistent with the echo.}\BibitemShut {Stop}%
\bibitem [{\citenamefont {Myers}\ \emph {et~al.}(2014)\citenamefont {Myers},
  \citenamefont {Das}, \citenamefont {Dartiailh}, \citenamefont {Ohno},
  \citenamefont {Awschalom},\ and\ \citenamefont
  {Bleszynski~Jayich}}]{Myers14}%
  \BibitemOpen
  \bibfield  {author} {\bibinfo {author} {\bibfnamefont {B.~A.}\ \bibnamefont
  {Myers}}, \bibinfo {author} {\bibfnamefont {A.}~\bibnamefont {Das}}, \bibinfo
  {author} {\bibfnamefont {M.~C.}\ \bibnamefont {Dartiailh}}, \bibinfo {author}
  {\bibfnamefont {K.}~\bibnamefont {Ohno}}, \bibinfo {author} {\bibfnamefont
  {D.~D.}\ \bibnamefont {Awschalom}},\ and\ \bibinfo {author} {\bibfnamefont
  {A.~C.}\ \bibnamefont {Bleszynski~Jayich}},\ }\href
  {https://doi.org/10.1103/PhysRevLett.113.027602} {\bibfield  {journal}
  {\bibinfo  {journal} {Physical Review Letters}\ }\textbf {\bibinfo {volume}
  {113}},\ \bibinfo {pages} {027602} (\bibinfo {year} {2014})}\BibitemShut
  {NoStop}%
\bibitem [{\citenamefont {Romach}\ \emph {et~al.}(2015)\citenamefont {Romach},
  \citenamefont {M{\"u}ller}, \citenamefont {Unden}, \citenamefont {Rogers},
  \citenamefont {Isoda}, \citenamefont {Itoh}, \citenamefont {Markham},
  \citenamefont {Stacey}, \citenamefont {Meijer}, \citenamefont {Pezzagna},
  \citenamefont {Naydenov}, \citenamefont {McGuinness}, \citenamefont
  {{Bar-Gill}},\ and\ \citenamefont {Jelezko}}]{Romach15}%
  \BibitemOpen
  \bibfield  {author} {\bibinfo {author} {\bibfnamefont {Y.}~\bibnamefont
  {Romach}}, \bibinfo {author} {\bibfnamefont {C.}~\bibnamefont {M{\"u}ller}},
  \bibinfo {author} {\bibfnamefont {T.}~\bibnamefont {Unden}}, \bibinfo
  {author} {\bibfnamefont {L.~J.}\ \bibnamefont {Rogers}}, \bibinfo {author}
  {\bibfnamefont {T.}~\bibnamefont {Isoda}}, \bibinfo {author} {\bibfnamefont
  {K.~M.}\ \bibnamefont {Itoh}}, \bibinfo {author} {\bibfnamefont
  {M.}~\bibnamefont {Markham}}, \bibinfo {author} {\bibfnamefont
  {A.}~\bibnamefont {Stacey}}, \bibinfo {author} {\bibfnamefont
  {J.}~\bibnamefont {Meijer}}, \bibinfo {author} {\bibfnamefont
  {S.}~\bibnamefont {Pezzagna}}, \bibinfo {author} {\bibfnamefont
  {B.}~\bibnamefont {Naydenov}}, \bibinfo {author} {\bibfnamefont {L.~P.}\
  \bibnamefont {McGuinness}}, \bibinfo {author} {\bibfnamefont
  {N.}~\bibnamefont {{Bar-Gill}}},\ and\ \bibinfo {author} {\bibfnamefont
  {F.}~\bibnamefont {Jelezko}},\ }\href
  {https://doi.org/10.1103/PhysRevLett.114.017601} {\bibfield  {journal}
  {\bibinfo  {journal} {Physical Review Letters}\ }\textbf {\bibinfo {volume}
  {114}},\ \bibinfo {pages} {017601} (\bibinfo {year} {2015})}\BibitemShut
  {NoStop}%
\bibitem [{Note2()}]{Note2}%
  \BibitemOpen
  \bibinfo {note} {We remark that while $S_w$ was initially introduced as it is
  consistent with R-E dynamics, additional $S_{\protect \text {\relax \protect
  \fontsize {5}{6}\protect \selectfont CP}}(\omega _m)$ independently reveal a
  nonvanishing baseline around $\protect \qopname \relax m{min}_m[S_{\protect
  \text {\relax \protect \fontsize {5}{6}\protect \selectfont CP}}(\omega
  _m)]=5(1)\protect \text {~ms}^{-1}$ [Fig.~\ref {FigSCNoiseModel}(a)]---in
  good agreement with $S_w$ predicted from an order of magnitude away with
  minimal experiment cost.}\BibitemShut {Stop}%
\bibitem [{\citenamefont {Stoffer}(1991)}]{Stoffer91}%
  \BibitemOpen
  \bibfield  {author} {\bibinfo {author} {\bibfnamefont {D.~S.}\ \bibnamefont
  {Stoffer}},\ }\href {https://doi.org/10.2307/2290595} {\bibfield  {journal}
  {\bibinfo  {journal} {Journal of the American Statistical Association}\
  }\textbf {\bibinfo {volume} {86}},\ \bibinfo {pages} {461} (\bibinfo {year}
  {1991})}\BibitemShut {NoStop}%
\bibitem [{\citenamefont {Hayes}\ \emph {et~al.}(2011)\citenamefont {Hayes},
  \citenamefont {Khodjasteh}, \citenamefont {Viola},\ and\ \citenamefont
  {Biercuk}}]{Hayes11}%
  \BibitemOpen
  \bibfield  {author} {\bibinfo {author} {\bibfnamefont {D.}~\bibnamefont
  {Hayes}}, \bibinfo {author} {\bibfnamefont {K.}~\bibnamefont {Khodjasteh}},
  \bibinfo {author} {\bibfnamefont {L.}~\bibnamefont {Viola}},\ and\ \bibinfo
  {author} {\bibfnamefont {M.~J.}\ \bibnamefont {Biercuk}},\ }\href
  {https://doi.org/10.1103/PhysRevA.84.062323} {\bibfield  {journal} {\bibinfo
  {journal} {Physical Review A}\ }\textbf {\bibinfo {volume} {84}},\ \bibinfo
  {pages} {062323} (\bibinfo {year} {2011})}\BibitemShut {NoStop}%
\bibitem [{\citenamefont {Cooper}\ \emph {et~al.}(2014)\citenamefont {Cooper},
  \citenamefont {Magesan}, \citenamefont {Yum},\ and\ \citenamefont
  {Cappellaro}}]{Cooper14}%
  \BibitemOpen
  \bibfield  {author} {\bibinfo {author} {\bibfnamefont {A.}~\bibnamefont
  {Cooper}}, \bibinfo {author} {\bibfnamefont {E.}~\bibnamefont {Magesan}},
  \bibinfo {author} {\bibfnamefont {H.~N.}\ \bibnamefont {Yum}},\ and\ \bibinfo
  {author} {\bibfnamefont {P.}~\bibnamefont {Cappellaro}},\ }\bibfield
  {journal} {\bibinfo  {journal} {Nature Communications}\ }\textbf {\bibinfo
  {volume} {5}},\ \href {https://doi.org/10.1038/ncomms4141}
  {10.1038/ncomms4141} (\bibinfo {year} {2014})\BibitemShut {NoStop}%
\bibitem [{\citenamefont {Cooper}\ \emph {et~al.}(2019)\citenamefont {Cooper},
  \citenamefont {Sun}, \citenamefont {Jaskula},\ and\ \citenamefont
  {Cappellaro}}]{Cooper19}%
  \BibitemOpen
  \bibfield  {author} {\bibinfo {author} {\bibfnamefont {A.}~\bibnamefont
  {Cooper}}, \bibinfo {author} {\bibfnamefont {W.~K.~C.}\ \bibnamefont {Sun}},
  \bibinfo {author} {\bibfnamefont {J.-C.}\ \bibnamefont {Jaskula}},\ and\
  \bibinfo {author} {\bibfnamefont {P.}~\bibnamefont {Cappellaro}},\ }\href
  {https://doi.org/10.1103/PhysRevApplied.12.044047} {\bibfield  {journal}
  {\bibinfo  {journal} {Physical Review Applied}\ }\textbf {\bibinfo {volume}
  {12}},\ \bibinfo {pages} {044047} (\bibinfo {year} {2019})}\BibitemShut
  {NoStop}%
\bibitem [{\citenamefont {Sun}\ \emph {et~al.}(2020)\citenamefont {Sun},
  \citenamefont {Cooper},\ and\ \citenamefont {Cappellaro}}]{Sun20}%
  \BibitemOpen
  \bibfield  {author} {\bibinfo {author} {\bibfnamefont {W.~K.~C.}\
  \bibnamefont {Sun}}, \bibinfo {author} {\bibfnamefont {A.}~\bibnamefont
  {Cooper}},\ and\ \bibinfo {author} {\bibfnamefont {P.}~\bibnamefont
  {Cappellaro}},\ }\href {https://doi.org/10.1103/PhysRevA.101.012319}
  {\bibfield  {journal} {\bibinfo  {journal} {Physical Review A}\ }\textbf
  {\bibinfo {volume} {101}},\ \bibinfo {pages} {012319} (\bibinfo {year}
  {2020})}\BibitemShut {NoStop}%
\bibitem [{\citenamefont {Belthangady}\ \emph {et~al.}(2013)\citenamefont
  {Belthangady}, \citenamefont {{Bar-Gill}}, \citenamefont {Pham},
  \citenamefont {Arai}, \citenamefont {Le~Sage}, \citenamefont {Cappellaro},\
  and\ \citenamefont {Walsworth}}]{Belthangady13}%
  \BibitemOpen
  \bibfield  {author} {\bibinfo {author} {\bibfnamefont {C.}~\bibnamefont
  {Belthangady}}, \bibinfo {author} {\bibfnamefont {N.}~\bibnamefont
  {{Bar-Gill}}}, \bibinfo {author} {\bibfnamefont {L.~M.}\ \bibnamefont
  {Pham}}, \bibinfo {author} {\bibfnamefont {K.}~\bibnamefont {Arai}}, \bibinfo
  {author} {\bibfnamefont {D.}~\bibnamefont {Le~Sage}}, \bibinfo {author}
  {\bibfnamefont {P.}~\bibnamefont {Cappellaro}},\ and\ \bibinfo {author}
  {\bibfnamefont {R.~L.}\ \bibnamefont {Walsworth}},\ }\href
  {https://doi.org/10.1103/PhysRevLett.110.157601} {\bibfield  {journal}
  {\bibinfo  {journal} {Physical Review Letters}\ }\textbf {\bibinfo {volume}
  {110}},\ \bibinfo {pages} {157601} (\bibinfo {year} {2013})}\BibitemShut
  {NoStop}%
\bibitem [{\citenamefont {Laraoui}\ and\ \citenamefont
  {Meriles}(2013)}]{Laraoui13}%
  \BibitemOpen
  \bibfield  {author} {\bibinfo {author} {\bibfnamefont {A.}~\bibnamefont
  {Laraoui}}\ and\ \bibinfo {author} {\bibfnamefont {C.~A.}\ \bibnamefont
  {Meriles}},\ }\href {https://doi.org/10.1021/nn400239n} {\bibfield  {journal}
  {\bibinfo  {journal} {ACS Nano}\ }\textbf {\bibinfo {volume} {7}},\ \bibinfo
  {pages} {3403} (\bibinfo {year} {2013})}\BibitemShut {NoStop}%
\bibitem [{\citenamefont {Sushkov}\ \emph {et~al.}(2014)\citenamefont
  {Sushkov}, \citenamefont {Lovchinsky}, \citenamefont {Chisholm},
  \citenamefont {Walsworth}, \citenamefont {Park},\ and\ \citenamefont
  {Lukin}}]{Sushkov14}%
  \BibitemOpen
  \bibfield  {author} {\bibinfo {author} {\bibfnamefont {A.~O.}\ \bibnamefont
  {Sushkov}}, \bibinfo {author} {\bibfnamefont {I.}~\bibnamefont {Lovchinsky}},
  \bibinfo {author} {\bibfnamefont {N.}~\bibnamefont {Chisholm}}, \bibinfo
  {author} {\bibfnamefont {R.~L.}\ \bibnamefont {Walsworth}}, \bibinfo {author}
  {\bibfnamefont {H.}~\bibnamefont {Park}},\ and\ \bibinfo {author}
  {\bibfnamefont {M.~D.}\ \bibnamefont {Lukin}},\ }\href
  {https://doi.org/10.1103/PhysRevLett.113.197601} {\bibfield  {journal}
  {\bibinfo  {journal} {Physical Review Letters}\ }\textbf {\bibinfo {volume}
  {113}},\ \bibinfo {pages} {197601} (\bibinfo {year} {2014})}\BibitemShut
  {NoStop}%
\bibitem [{\citenamefont {Knowles}\ \emph {et~al.}(2016)\citenamefont
  {Knowles}, \citenamefont {Kara},\ and\ \citenamefont
  {Atat{\"u}re}}]{Knowles16}%
  \BibitemOpen
  \bibfield  {author} {\bibinfo {author} {\bibfnamefont {H.~S.}\ \bibnamefont
  {Knowles}}, \bibinfo {author} {\bibfnamefont {D.~M.}\ \bibnamefont {Kara}},\
  and\ \bibinfo {author} {\bibfnamefont {M.}~\bibnamefont {Atat{\"u}re}},\
  }\href {https://doi.org/10.1103/PhysRevLett.117.100802} {\bibfield  {journal}
  {\bibinfo  {journal} {Physical Review Letters}\ }\textbf {\bibinfo {volume}
  {117}},\ \bibinfo {pages} {100802} (\bibinfo {year} {2016})}\BibitemShut
  {NoStop}%
\bibitem [{\citenamefont {Rosenfeld}\ \emph {et~al.}(2018)\citenamefont
  {Rosenfeld}, \citenamefont {Pham}, \citenamefont {Lukin},\ and\ \citenamefont
  {Walsworth}}]{Rosenfeld18}%
  \BibitemOpen
  \bibfield  {author} {\bibinfo {author} {\bibfnamefont {E.~L.}\ \bibnamefont
  {Rosenfeld}}, \bibinfo {author} {\bibfnamefont {L.~M.}\ \bibnamefont {Pham}},
  \bibinfo {author} {\bibfnamefont {M.~D.}\ \bibnamefont {Lukin}},\ and\
  \bibinfo {author} {\bibfnamefont {R.~L.}\ \bibnamefont {Walsworth}},\ }\href
  {https://doi.org/10.1103/PhysRevLett.120.243604} {\bibfield  {journal}
  {\bibinfo  {journal} {Physical Review Letters}\ }\textbf {\bibinfo {volume}
  {120}},\ \bibinfo {pages} {243604} (\bibinfo {year} {2018})}\BibitemShut
  {NoStop}%
\bibitem [{\citenamefont {Pinto}\ \emph {et~al.}(2020)\citenamefont {Pinto},
  \citenamefont {Paone}, \citenamefont {Kern}, \citenamefont {Dierker},
  \citenamefont {Wieczorek}, \citenamefont {Singha}, \citenamefont {Dasari},
  \citenamefont {Finkler}, \citenamefont {Harneit}, \citenamefont {Wrachtrup},\
  and\ \citenamefont {Kern}}]{Pinto20}%
  \BibitemOpen
  \bibfield  {author} {\bibinfo {author} {\bibfnamefont {D.}~\bibnamefont
  {Pinto}}, \bibinfo {author} {\bibfnamefont {D.}~\bibnamefont {Paone}},
  \bibinfo {author} {\bibfnamefont {B.}~\bibnamefont {Kern}}, \bibinfo {author}
  {\bibfnamefont {T.}~\bibnamefont {Dierker}}, \bibinfo {author} {\bibfnamefont
  {R.}~\bibnamefont {Wieczorek}}, \bibinfo {author} {\bibfnamefont
  {A.}~\bibnamefont {Singha}}, \bibinfo {author} {\bibfnamefont
  {D.}~\bibnamefont {Dasari}}, \bibinfo {author} {\bibfnamefont
  {A.}~\bibnamefont {Finkler}}, \bibinfo {author} {\bibfnamefont
  {W.}~\bibnamefont {Harneit}}, \bibinfo {author} {\bibfnamefont
  {J.}~\bibnamefont {Wrachtrup}},\ and\ \bibinfo {author} {\bibfnamefont
  {K.}~\bibnamefont {Kern}},\ }\href
  {https://doi.org/10.1038/s41467-020-20202-3} {\bibfield  {journal} {\bibinfo
  {journal} {Nature Communications}\ }\textbf {\bibinfo {volume} {11}},\
  \bibinfo {pages} {6405} (\bibinfo {year} {2020})}\BibitemShut {NoStop}%
\bibitem [{\citenamefont {Degen}\ \emph {et~al.}(2021)\citenamefont {Degen},
  \citenamefont {Loenen}, \citenamefont {Bartling}, \citenamefont {Bradley},
  \citenamefont {Meinsma}, \citenamefont {Markham}, \citenamefont {Twitchen},\
  and\ \citenamefont {Taminiau}}]{Degen21}%
  \BibitemOpen
  \bibfield  {author} {\bibinfo {author} {\bibfnamefont {M.~J.}\ \bibnamefont
  {Degen}}, \bibinfo {author} {\bibfnamefont {S.~J.~H.}\ \bibnamefont
  {Loenen}}, \bibinfo {author} {\bibfnamefont {H.~P.}\ \bibnamefont
  {Bartling}}, \bibinfo {author} {\bibfnamefont {C.~E.}\ \bibnamefont
  {Bradley}}, \bibinfo {author} {\bibfnamefont {A.~L.}\ \bibnamefont
  {Meinsma}}, \bibinfo {author} {\bibfnamefont {M.}~\bibnamefont {Markham}},
  \bibinfo {author} {\bibfnamefont {D.~J.}\ \bibnamefont {Twitchen}},\ and\
  \bibinfo {author} {\bibfnamefont {T.~H.}\ \bibnamefont {Taminiau}},\ }\href
  {https://doi.org/10.1038/s41467-021-23454-9} {\bibfield  {journal} {\bibinfo
  {journal} {Nature Communications}\ }\textbf {\bibinfo {volume} {12}},\
  \bibinfo {pages} {3470} (\bibinfo {year} {2021})}\BibitemShut {NoStop}%
\bibitem [{\citenamefont {Kucsko}\ \emph {et~al.}(2018)\citenamefont {Kucsko},
  \citenamefont {Choi}, \citenamefont {Choi}, \citenamefont {Maurer},
  \citenamefont {Zhou}, \citenamefont {Landig}, \citenamefont {Sumiya},
  \citenamefont {Onoda}, \citenamefont {Isoya}, \citenamefont {Jelezko},
  \citenamefont {Demler}, \citenamefont {Yao},\ and\ \citenamefont
  {Lukin}}]{Kucsko18}%
  \BibitemOpen
  \bibfield  {author} {\bibinfo {author} {\bibfnamefont {G.}~\bibnamefont
  {Kucsko}}, \bibinfo {author} {\bibfnamefont {S.}~\bibnamefont {Choi}},
  \bibinfo {author} {\bibfnamefont {J.}~\bibnamefont {Choi}}, \bibinfo {author}
  {\bibfnamefont {P.~C.}\ \bibnamefont {Maurer}}, \bibinfo {author}
  {\bibfnamefont {H.}~\bibnamefont {Zhou}}, \bibinfo {author} {\bibfnamefont
  {R.}~\bibnamefont {Landig}}, \bibinfo {author} {\bibfnamefont
  {H.}~\bibnamefont {Sumiya}}, \bibinfo {author} {\bibfnamefont
  {S.}~\bibnamefont {Onoda}}, \bibinfo {author} {\bibfnamefont
  {J.}~\bibnamefont {Isoya}}, \bibinfo {author} {\bibfnamefont
  {F.}~\bibnamefont {Jelezko}}, \bibinfo {author} {\bibfnamefont
  {E.}~\bibnamefont {Demler}}, \bibinfo {author} {\bibfnamefont {N.~Y.}\
  \bibnamefont {Yao}},\ and\ \bibinfo {author} {\bibfnamefont {M.~D.}\
  \bibnamefont {Lukin}},\ }\href
  {https://doi.org/10.1103/PhysRevLett.121.023601} {\bibfield  {journal}
  {\bibinfo  {journal} {Physical Review Letters}\ }\textbf {\bibinfo {volume}
  {121}},\ \bibinfo {pages} {023601} (\bibinfo {year} {2018})}\BibitemShut
  {NoStop}%
\bibitem [{\citenamefont {Klauder}\ and\ \citenamefont
  {Anderson}(1962)}]{Klauder62}%
  \BibitemOpen
  \bibfield  {author} {\bibinfo {author} {\bibfnamefont {J.~R.}\ \bibnamefont
  {Klauder}}\ and\ \bibinfo {author} {\bibfnamefont {P.~W.}\ \bibnamefont
  {Anderson}},\ }\href {https://doi.org/10.1103/PhysRev.125.912} {\bibfield
  {journal} {\bibinfo  {journal} {Physical Review}\ }\textbf {\bibinfo {volume}
  {125}},\ \bibinfo {pages} {912} (\bibinfo {year} {1962})}\BibitemShut
  {NoStop}%
\bibitem [{\citenamefont {Dobrovitski}\ \emph {et~al.}(2009)\citenamefont
  {Dobrovitski}, \citenamefont {Feiguin}, \citenamefont {Hanson},\ and\
  \citenamefont {Awschalom}}]{Dobrovitski09}%
  \BibitemOpen
  \bibfield  {author} {\bibinfo {author} {\bibfnamefont {V.~V.}\ \bibnamefont
  {Dobrovitski}}, \bibinfo {author} {\bibfnamefont {A.~E.}\ \bibnamefont
  {Feiguin}}, \bibinfo {author} {\bibfnamefont {R.}~\bibnamefont {Hanson}},\
  and\ \bibinfo {author} {\bibfnamefont {D.~D.}\ \bibnamefont {Awschalom}},\
  }\href {https://doi.org/10.1103/PhysRevLett.102.237601} {\bibfield  {journal}
  {\bibinfo  {journal} {Physical Review Letters}\ }\textbf {\bibinfo {volume}
  {102}},\ \bibinfo {pages} {237601} (\bibinfo {year} {2009})}\BibitemShut
  {NoStop}%
\bibitem [{\citenamefont {Bauch}\ \emph {et~al.}(2020)\citenamefont {Bauch},
  \citenamefont {Singh}, \citenamefont {Lee}, \citenamefont {Hart},
  \citenamefont {Schloss}, \citenamefont {Turner}, \citenamefont {Barry},
  \citenamefont {Pham}, \citenamefont {{Bar-Gill}}, \citenamefont {Yelin},\
  and\ \citenamefont {Walsworth}}]{Bauch20}%
  \BibitemOpen
  \bibfield  {author} {\bibinfo {author} {\bibfnamefont {E.}~\bibnamefont
  {Bauch}}, \bibinfo {author} {\bibfnamefont {S.}~\bibnamefont {Singh}},
  \bibinfo {author} {\bibfnamefont {J.}~\bibnamefont {Lee}}, \bibinfo {author}
  {\bibfnamefont {C.~A.}\ \bibnamefont {Hart}}, \bibinfo {author}
  {\bibfnamefont {J.~M.}\ \bibnamefont {Schloss}}, \bibinfo {author}
  {\bibfnamefont {M.~J.}\ \bibnamefont {Turner}}, \bibinfo {author}
  {\bibfnamefont {J.~F.}\ \bibnamefont {Barry}}, \bibinfo {author}
  {\bibfnamefont {L.~M.}\ \bibnamefont {Pham}}, \bibinfo {author}
  {\bibfnamefont {N.}~\bibnamefont {{Bar-Gill}}}, \bibinfo {author}
  {\bibfnamefont {S.~F.}\ \bibnamefont {Yelin}},\ and\ \bibinfo {author}
  {\bibfnamefont {R.~L.}\ \bibnamefont {Walsworth}},\ }\href
  {https://doi.org/10.1103/PhysRevB.102.134210} {\bibfield  {journal} {\bibinfo
   {journal} {Physical Review B}\ }\textbf {\bibinfo {volume} {102}},\ \bibinfo
  {pages} {134210} (\bibinfo {year} {2020})}\BibitemShut {NoStop}%
\bibitem [{\citenamefont {Grinolds}\ \emph {et~al.}(2013)\citenamefont
  {Grinolds}, \citenamefont {Hong}, \citenamefont {Maletinsky}, \citenamefont
  {Luan}, \citenamefont {Lukin}, \citenamefont {Walsworth},\ and\ \citenamefont
  {Yacoby}}]{Grinolds13}%
  \BibitemOpen
  \bibfield  {author} {\bibinfo {author} {\bibfnamefont {M.~S.}\ \bibnamefont
  {Grinolds}}, \bibinfo {author} {\bibfnamefont {S.}~\bibnamefont {Hong}},
  \bibinfo {author} {\bibfnamefont {P.}~\bibnamefont {Maletinsky}}, \bibinfo
  {author} {\bibfnamefont {L.}~\bibnamefont {Luan}}, \bibinfo {author}
  {\bibfnamefont {M.~D.}\ \bibnamefont {Lukin}}, \bibinfo {author}
  {\bibfnamefont {R.~L.}\ \bibnamefont {Walsworth}},\ and\ \bibinfo {author}
  {\bibfnamefont {A.}~\bibnamefont {Yacoby}},\ }\href
  {https://doi.org/10.1038/nphys2543} {\bibfield  {journal} {\bibinfo
  {journal} {Nature Physics}\ }\textbf {\bibinfo {volume} {9}},\ \bibinfo
  {pages} {215} (\bibinfo {year} {2013})}\BibitemShut {NoStop}%
\bibitem [{\citenamefont {Sza{\'n}kowski}\ \emph {et~al.}(2016)\citenamefont
  {Sza{\'n}kowski}, \citenamefont {Trippenbach},\ and\ \citenamefont
  {Cywi{\'n}ski}}]{Szankowski16}%
  \BibitemOpen
  \bibfield  {author} {\bibinfo {author} {\bibfnamefont {P.}~\bibnamefont
  {Sza{\'n}kowski}}, \bibinfo {author} {\bibfnamefont {M.}~\bibnamefont
  {Trippenbach}},\ and\ \bibinfo {author} {\bibfnamefont {{\L}.}~\bibnamefont
  {Cywi{\'n}ski}},\ }\href {https://doi.org/10.1103/PhysRevA.94.012109}
  {\bibfield  {journal} {\bibinfo  {journal} {Physical Review A}\ }\textbf
  {\bibinfo {volume} {94}},\ \bibinfo {pages} {012109} (\bibinfo {year}
  {2016})}\BibitemShut {NoStop}%
\bibitem [{\citenamefont {{Paz-Silva}}\ \emph {et~al.}(2017)\citenamefont
  {{Paz-Silva}}, \citenamefont {Norris},\ and\ \citenamefont
  {Viola}}]{Paz-Silva17}%
  \BibitemOpen
  \bibfield  {author} {\bibinfo {author} {\bibfnamefont {G.~A.}\ \bibnamefont
  {{Paz-Silva}}}, \bibinfo {author} {\bibfnamefont {L.~M.}\ \bibnamefont
  {Norris}},\ and\ \bibinfo {author} {\bibfnamefont {L.}~\bibnamefont
  {Viola}},\ }\href {https://doi.org/10.1103/PhysRevA.95.022121} {\bibfield
  {journal} {\bibinfo  {journal} {Physical Review A}\ }\textbf {\bibinfo
  {volume} {95}},\ \bibinfo {pages} {022121} (\bibinfo {year}
  {2017})}\BibitemShut {NoStop}%
\bibitem [{\citenamefont {{von L{\"u}pke}}\ \emph {et~al.}(2020)\citenamefont
  {{von L{\"u}pke}}, \citenamefont {Beaudoin}, \citenamefont {Norris},
  \citenamefont {Sung}, \citenamefont {Winik}, \citenamefont {Qiu},
  \citenamefont {Kjaergaard}, \citenamefont {Kim}, \citenamefont {Yoder},
  \citenamefont {Gustavsson}, \citenamefont {Viola},\ and\ \citenamefont
  {Oliver}}]{vonLupke20}%
  \BibitemOpen
  \bibfield  {author} {\bibinfo {author} {\bibfnamefont {U.}~\bibnamefont {{von
  L{\"u}pke}}}, \bibinfo {author} {\bibfnamefont {F.}~\bibnamefont {Beaudoin}},
  \bibinfo {author} {\bibfnamefont {L.~M.}\ \bibnamefont {Norris}}, \bibinfo
  {author} {\bibfnamefont {Y.}~\bibnamefont {Sung}}, \bibinfo {author}
  {\bibfnamefont {R.}~\bibnamefont {Winik}}, \bibinfo {author} {\bibfnamefont
  {J.~Y.}\ \bibnamefont {Qiu}}, \bibinfo {author} {\bibfnamefont
  {M.}~\bibnamefont {Kjaergaard}}, \bibinfo {author} {\bibfnamefont
  {D.}~\bibnamefont {Kim}}, \bibinfo {author} {\bibfnamefont {J.}~\bibnamefont
  {Yoder}}, \bibinfo {author} {\bibfnamefont {S.}~\bibnamefont {Gustavsson}},
  \bibinfo {author} {\bibfnamefont {L.}~\bibnamefont {Viola}},\ and\ \bibinfo
  {author} {\bibfnamefont {W.~D.}\ \bibnamefont {Oliver}},\ }\href
  {https://doi.org/10.1103/PRXQuantum.1.010305} {\bibfield  {journal} {\bibinfo
   {journal} {PRX Quantum}\ }\textbf {\bibinfo {volume} {1}},\ \bibinfo {pages}
  {010305} (\bibinfo {year} {2020})}\BibitemShut {NoStop}%
\bibitem [{\citenamefont {Layden}\ \emph {et~al.}(2020)\citenamefont {Layden},
  \citenamefont {Chen},\ and\ \citenamefont {Cappellaro}}]{Layden20}%
  \BibitemOpen
  \bibfield  {author} {\bibinfo {author} {\bibfnamefont {D.}~\bibnamefont
  {Layden}}, \bibinfo {author} {\bibfnamefont {M.}~\bibnamefont {Chen}},\ and\
  \bibinfo {author} {\bibfnamefont {P.}~\bibnamefont {Cappellaro}},\ }\href
  {https://doi.org/10.1103/PhysRevLett.124.020504} {\bibfield  {journal}
  {\bibinfo  {journal} {Physical Review Letters}\ }\textbf {\bibinfo {volume}
  {124}},\ \bibinfo {pages} {020504} (\bibinfo {year} {2020})}\BibitemShut
  {NoStop}%
\bibitem [{\citenamefont {Layden}\ and\ \citenamefont
  {Cappellaro}(2018)}]{Layden18b}%
  \BibitemOpen
  \bibfield  {author} {\bibinfo {author} {\bibfnamefont {D.}~\bibnamefont
  {Layden}}\ and\ \bibinfo {author} {\bibfnamefont {P.}~\bibnamefont
  {Cappellaro}},\ }\href {https://doi.org/10.1038/s41534-018-0082-2} {\bibfield
   {journal} {\bibinfo  {journal} {npj Quantum Information}\ }\textbf {\bibinfo
  {volume} {4}},\ \bibinfo {pages} {1} (\bibinfo {year} {2018})}\BibitemShut
  {NoStop}%
\bibitem [{\citenamefont {Abragam}(1961)}]{Abragam61}%
  \BibitemOpen
  \bibfield  {author} {\bibinfo {author} {\bibfnamefont {A.}~\bibnamefont
  {Abragam}},\ }\href@noop {} {\emph {\bibinfo {title} {The {{Principles}} of
  {{Nuclear Magnetism}}}}}\ (\bibinfo  {publisher} {{Clarendon Press}},\
  \bibinfo {year} {1961})\BibitemShut {NoStop}%
\bibitem [{\citenamefont {Hoch}\ and\ \citenamefont
  {Reynhardt}(1988)}]{Hoch88}%
  \BibitemOpen
  \bibfield  {author} {\bibinfo {author} {\bibfnamefont {M.~J.~R.}\
  \bibnamefont {Hoch}}\ and\ \bibinfo {author} {\bibfnamefont {E.~C.}\
  \bibnamefont {Reynhardt}},\ }\href {https://doi.org/10.1103/PhysRevB.37.9222}
  {\bibfield  {journal} {\bibinfo  {journal} {Physical Review B}\ }\textbf
  {\bibinfo {volume} {37}},\ \bibinfo {pages} {9222} (\bibinfo {year}
  {1988})}\BibitemShut {NoStop}%
\bibitem [{\citenamefont {{Bar-Gill}}\ \emph {et~al.}(2012)\citenamefont
  {{Bar-Gill}}, \citenamefont {Pham}, \citenamefont {Belthangady},
  \citenamefont {Le~Sage}, \citenamefont {Cappellaro}, \citenamefont {Maze},
  \citenamefont {Lukin}, \citenamefont {Yacoby},\ and\ \citenamefont
  {Walsworth}}]{Bar-Gill12}%
  \BibitemOpen
  \bibfield  {author} {\bibinfo {author} {\bibfnamefont {N.}~\bibnamefont
  {{Bar-Gill}}}, \bibinfo {author} {\bibfnamefont {L.~M.}\ \bibnamefont
  {Pham}}, \bibinfo {author} {\bibfnamefont {C.}~\bibnamefont {Belthangady}},
  \bibinfo {author} {\bibfnamefont {D.}~\bibnamefont {Le~Sage}}, \bibinfo
  {author} {\bibfnamefont {P.}~\bibnamefont {Cappellaro}}, \bibinfo {author}
  {\bibfnamefont {J.~R.}\ \bibnamefont {Maze}}, \bibinfo {author}
  {\bibfnamefont {M.~D.}\ \bibnamefont {Lukin}}, \bibinfo {author}
  {\bibfnamefont {A.}~\bibnamefont {Yacoby}},\ and\ \bibinfo {author}
  {\bibfnamefont {R.}~\bibnamefont {Walsworth}},\ }\href
  {https://doi.org/10.1038/ncomms1856} {\bibfield  {journal} {\bibinfo
  {journal} {Nature Communications}\ }\textbf {\bibinfo {volume} {3}},\
  \bibinfo {pages} {858} (\bibinfo {year} {2012})}\BibitemShut {NoStop}%
\end{thebibliography}%

\end{document}